\documentclass[fleqn,usenatbib]{mnras}
\usepackage{newtxtext,newtxmath}

\usepackage[T1]{fontenc}
\usepackage{ae,aecompl}
\usepackage{graphicx,epsfig}	
\usepackage{amsmath}	
\usepackage{booktabs} 
\usepackage[dvipsnames]{xcolor}
\usepackage{url}

\newcommand{\msun}{M$\mathrm{_\odot}$}

\newcommand{\mbh}{$M{\mathrm{_{BH}}}$\,}

\newcommand{\Mstar}{$M_{*}$\,}

\newcommand{\kms}{km~s${-1}$}

\newcommand{\re}{R_{\rm e}}

\title[AGNs in low-mass compact galaxies]{A search for active galactic nuclei in low-mass compact galaxies}

\author[Ferr\'e-Mateu et  al.]{A. Ferr\'e-Mateu$^{1,2}$\thanks{E-mail: aferremateu@icc.ub.edu}, M. Mezcua$^{3,4}$\thanks{E-mail: marmezcua.astro@gmail.com}, R. S. Barrows$^{5}$ \\
$^{1}$ Institut de Ciencies del Cosmos (ICCUB), Universitat de Barcelona (IEEC-UB), E02028 Barcelona, Spain\\
$^{2}$ Centre for Astrophysics \& Supercomputing, Swinburne University of Technology, Hawthorn VIC 3122, Australia\\
$^{3}$ Institute of Space Sciences (ICE, CSIC), Campus UAB, Carrer de Magrans, 08193 Barcelona, Spain\\
$^{4}$ Institut d'Estudis Espacials de Catalunya (IEEC), Carrer Gran Capit\`a, 08034 Barcelona, Spain\\
$^{5}$ Department of Astrophysical and Planetary Sciences, University of Colorado Boulder, Boulder, CO 80309, USA\\
}

\date{Accepted 2021 July 2. Received 2021 July 2; in original form 2021 April 10}

\pubyear{2021}

\begin{document}
\label{firstpage}
\pagerange{\pageref{firstpage}--\pageref{lastpage}}
\maketitle

\begin{abstract}
Low-mass compact galaxies (ultracompact dwarfs -UCDs- and compact ellipticals -cEs-) populate the stellar size-mass plane between globular clusters and early-type galaxies. Known to be formed either in-situ with an intrinsically low mass or resulting from the stripping of a more massive galaxy, the presence of a supermassive or an intermediate-mass black hole (BH) could help discriminate between these possible scenarios. With this aim, we have performed a multiwavelength search of active BH activity, i.e. active galactic nuclei (AGN), in a sample of 937 low-mass compact galaxies (580 UCDs and 357 cEs). This constitutes the largest study of AGN activity in these types of galaxies. Based on their X-ray luminosity, radio luminosity and morphology, and/or optical emission line diagnostic diagrams, we find a total of 11 cEs that host an AGN. We also study for the first time the location of both low-mass compact galaxies (UCDs and cEs) and dwarf galaxies hosting AGN on the BH-galaxy scaling relations, finding that low-mass compact galaxies tend to be overmassive in the BH mass-stellar mass plane but not as much in the BH mass-stellar velocity dispersion correlation. This, together with available BH mass measurements for some of the low-mass compact galaxies, supports a stripping origin for the majority of these objects that would contribute to the scatter seen at the low-mass end of the BH-galaxy scaling relations. However, the differences are too large to be explained solely by this scatter, and thus our results suggest that a flattening at such low-masses is also plausible, happening at a velocity dispersion of $\sim$20-40 \kms.
\end{abstract}


\section{Introduction}
The low-mass end of the realm of early-type galaxies (ETGs) has been populated in the recent years by a series of compact objects found to bridge the gap between globular clusters and massive ETGs (e.g. \citealt{Brodie2011}; \citealt{Misgeld2011}). With characteristic stellar sizes and stellar masses (\Mstar) defining the transitions from one type to the other, the origin of these families are still strongly debated.

From one side, there are the lowest-mass and most compact galaxies, the so-called ultra compact dwarfs (UCDs), which range from 10$^{6}$ \la \Mstar(\msun) $\la 10^{8}$ and have sizes with effective radii ($\re$) between 10 and 100pc. Following an intrinsic formation channel, UCDs have been proposed to simply be the massive end of the globular cluster family (e.g. \citealt{Mieske2012}) or to result from young massive star clusters being aggregated during violent gas-rich galaxy mergers (e.g. \citealt{Fellhauer2002}; \citealt{Bruns2011}; \citealt{Mahani2021}). However, UCDs could also be the leftovers of a larger and more massive galaxy that lost the majority of its stellar mass due to external processes, such as tidal or ram-pressure stripping (e.g. \citealt{Faber1973}; \citealt{Bekki2001}; \citealt{Drinkwater2003}). The progenitor galaxy is expected to be typically a dwarf elliptical or a low-mass ETG. The majority of these progenitor galaxies are expected to host a nuclear star cluster (NSC) in their centers. Therefore, what is left behind as UCD would be the threshed NSC (in the case of the low-mass UCDs, \Mstar$<10^{7}$\msun) while for more massive UCDs (\Mstar$>10^{7}$\msun) it would be NSC surrounded by a diffuse region that contains part of the progenitor galaxy remains (e.g. \citealt{Evstigneeva2007}; \citealt{Pfeffer2013}). 

Similar formation channels are also proposed for the slightly more massive counterparts, the compact ellipticals (cEs), coming in both flavours of intrinsic and external processes. cEs populate the regime 10$^{8}$ \la \Mstar(\msun) $\la 10^{10}$ and have sizes between 100 $\la \re$(pc) $\la$ 900. They can also be the result of the removal of stars from a more massive ETG, a spiral galaxy (e.g. \citealt{Bekki2003}; \citealt{Graham2003}) or from a compact massive galaxy \citep{Ferre-Mateu2021}. While being the remnant of a NSC+diffuse part is plausible at these higher stellar masses, the occupation fraction of NSC decreases with increasing stellar mass and by \Mstar$\ga10^{10}$\msun the object that dominates in the center of the galaxy will be a black hole (BH; e.g. \citealt{Ferrarese2006}; \citealt{Cote2006}). Therefore, in the case of cEs, whose progenitor tends to be more a massive galaxy, the remnant is more similar to the innermost part of high-redshift red nuggets (e.g. \citealt{Huang2013}; \citealt{Zolotov2015}); see Fig. 4 in \citealt{Ferre-Mateu2021} for a review on the formation channels of cEs). Alternatively, cEs could also be intrinsically low mass, compact objects that formed as we see them today (\citealt{Wirth1984}; \citealt{Kormendy2009, Kormendy2012}). 

Observational evidence for these different origins have been found, from compact galaxies caught in the act of being stripped (e.g. \citealt{Chilingarian2009}; \citealt{Huxor2011}; \citealt{Paudel2013}; \citealt{Chilingarian2015}; \citealt{Ferre-Mateu2018}) to the existence of them in isolated places that would make the stripping process impossible (e.g. \citealt{Paudel2014}; \citealt{Ferre-Mateu2018}; \citealt{Kim2020}). In terms of simulations, \citet{Bekki2001}, \citet{Pfeffer2013, Pfeffer2016} and \citet{Goodman2018} have reproduced the stripping path for UCDs, whereas \citet{Martinovic2017}, \citet{Du2019} and \citet{UrrutiaZapata2019} have shown that both being formed in-situ as low-mass galaxies but also being the result of stripping larger galaxies can reproduce the observed properties of cEs. Although the numbers of both families are still very limited, understanding their nature is vital to understand the complete baryonic budget of the Universe, and thus obtaining the proportion of in-situ formed (intrinsic) \textit{vs} external processes (stripping) objects in each family is a critical effort. Some works have already tried to address these different origins via the study of their kinematics, stellar populations and other relevant properties (e.g. \citealt{Chilingarian2008,Chilingarian2009}; \citealt{Norris2014}; \citealt{Janz2015}; \citealt{Chilingarian2015}; \citealt{Ferre-Mateu2018}; \citealt{Kim2020}; \citealt{Ferre-Mateu2021}). 

However, one characteristic still remains quite elusive to date: the existence or not of BHs at their centers. Typically, both UCDs and cEs show enhanced ratios of dynamical-to-stellar masses (e.g. \citealt{Chilingarian2008}; \citealt{Mieske2013}; \citealt{Forbes2014}; \citealt{Janz2015}; \citealt{Ferre-Mateu2018, Ferre-Mateu2021}). While this could be the effect of a different initial mass function (e.g. \citealt{Mieske2008}; \citealt{Ferre-Mateu2013}) or accounted by the presence of large dark matter halos (e.g. \citealt{Hasegan2007}; \citealt{Chilingarian2008}), the most plausible explanation is that they host a BH at their centers. If UCDs and cEs are formed intrinsically, it is expected that they will host BHs that follow the scaling relations: ranging from the elusive intermediate-mass BHs (IMBHs; \mbh$\sim 10^{2}$--10$^5$\msun) in the case of UCDs, up to the lower limit of super-massive BHs (SMBHs; \mbh$\gtrsim 10^{6}$\msun) for the cEs. Although the presence of IMBHs could also be expected as the result of runaway merging \citep{Shi2020}, it would not be expected if low-mass compact objects follow a stripping formation channel. In such cases, one would expect to find SMBHs of \mbh$\sim 10^{6}$--10$^9$\msun. As the progenitor galaxy is expected to follow the scaling relations governing massive galaxies up to the stripping event, the aftermath will be an untouched BH in the center. Therefore, these galaxies will appear as outliers in the local BH-galaxy scaling relations (e.g. \citealt{Ferre-Mateu2015}; \citealt{vanSon2019}). 

Tens of IMBH candidates have been found in dwarf galaxies as low-mass (\mbh$\lesssim10^{6}$\msun) active galactic nuclei (AGN); however, most of these dwarf galaxies are either star-forming or of late-type (e.g. \citealt{Jiang2011}; \citealt{Mezcua2016, Mezcua2018a, Mezcua2020};  \citealt{Chilingarian2018}; \citealt{Kimbrell2021}; see \citealt{Mezcua2017}; \citealt{Greene2020} for reviews). Therefore, the number of detected BHs in low-mass compact galaxies is extremely scarce. The majority of such detections have been reported in UCDs (e.g. \citealt{Seth2014}; \citealt{Ahn2017, Ahn2018}; \citealt{Afanasiev2018} and \citealt{Voggel2018,Voggel2019}), with the caveat that they are limited to UCDs with relatively high stellar masses ($>10^{7}$\msun\,) and therefore the measured BH masses measured to date are compatible with being all SMBHs (\mbh$>10^{6}$\msun). The number of cEs with reported BHs is even lower (e.g. M32, \citealt{vanderMarel1997}; J085431.18+173730.5, \citealt{Paudel2016}; or NGC\,741, \citealt{Schellenberger2017}). 

Motivated by these findings, we aim at searching for AGN signatures indicative of the presence of an active BHs at the centers of UCDs and cEs. This can help further discriminate between the possible origins for these galaxies and pose constraints on the formation channels of the objects in the compact low-mass end. We first search for X-ray signatures, following a similar approach as in \citet{Pandya2016} and \citet{Hou2016}. They both found a detection rate of $\sim$3$\%$, consistent in all cases with low-mass X-ray binaries. Here, we complement these works by extending the search in a larger sample of spectroscopically confirmed UCDs and include, for the first time, a number of known cEs. The search is thus done to a total sample of 937 objects, making this the largest dataset of low-mass compact galaxies to date (Section \S \ref{section:sample}). In addition to the search in X-rays, we also search for AGN signatures in the optical, infrared, and radio regimes (Section \S \ref{section:data}). The results obtained are presented in Section \S \ref{section:results}, and we discuss on the implications for the type of BHs in these objects in Section \S \ref{section:discussion}. Final conclusions are provided in Section \S \ref{section:conclusions}.

\section{Sample}\label{section:sample}
We compile the largest available sample of low-mass compact galaxies comprising fiducial UCDs and cEs that can be found in the literature, checking for duplicates within compilations. We have focused on those samples with spectroscopic identifications to confirm their galaxy type. Most of the sources included have also other relevant information available, such as the stellar mass and galaxy size. Although UCDs are less massive and more compact than cEs, they have been detected in larger numbers and therefore our sample is biased towards larger numbers of UCDs. 

This way, we build our sample by considering the UCDs from \citet{Pandya2016}, the largest compilation to date. While it consists of 578 UCDs, several of them are only candidates as they are not spectroscopically confirmed. For this work we select only the confirmed cases (227). We then add those in \citet{Fahrion2019} that are not found in common (137 UCDs more). We note that the sample of \citet{Hou2016} is not included here as there is no information separating the UCDs from the extended stellar clusters, but there is a large overlap in objects from the previous catalogues. We also include UCDs in the Virgo cluster from different works (\citealt{Liu2015}; \citealt{Zhang2018}; \citealt{Forbes2020}; 96 UCDs) and a sample of 20 UCDs around NGC\,3115 \citep{Dolfi2020}. We also consider those compact galaxies from the compilation of \citet{Norris2014}, which is a mixture of both UCDs (97) and cEs (39). We also add cEs from the Sloan Digital Sky Survey (SDSS) compilations, from both \citet{Chilingarian2015} and \citet{Kim2020} (195 and 101, respectively, private communication for the \citealt{Kim2020} catalogue). Furthermore, we include cEs in the Virgo cluster from \citet{Guerou2015} and other new cEs from \citet{Ferre-Mateu2018} and \citet{Ferre-Mateu2021}. Some other individual objects that have well studied properties have also been included in our search, such as NGC7727-Nuc2 \citep{Schweizer2018}, NGC5044-UCD1 \citep{Faifer2017}, and one cE which is a known AGN \citep{Paudel2016}. 

We have to warn here about a possible caveat regarding the \citet{Norris2014} sample, which is that it does not differentiate between globular clusters and UCDs. While globular clusters are typically selected as having log M$_{*}\lesssim$ 5 \msun, the fact that some UCDs are expected to be the massive tail of globular clusters results in the 6 $<$ log (\Mstar/\msun) $<$ 7 regime being a mixture of both types, in what is known as a transition zone (e.g. \citealt{Brodie2011}; \citealt{Norris2014}). Therefore, a stellar mass cut alone is not enough to separate them. A size cut is not recommended either as, although UCDs have been typically described as having a size $>$10\,pc, many have been found with smaller sizes (e.g. \citealt{Forbes2014}). Because of this, we apply a magnitude cut in addition to the \Mstar\ one following \citet{Mieske2012}, with UCDs having M$_\mathrm{V}<-$10.25. This ensures that no globular clusters are included in the sample. 

Table \ref{tab:numbers} summarizes the different works used to create our sample, with the number of galaxies of each type used. In total, we consider 937 low-mass compact galaxies (357 cEs and 580 UCDs). The sampled galaxies span out to z=0.113.\\

\begin{table}
\centering
\caption{Summary of literature low-mass compact galaxies used}
\label{tab:numbers}
\begin{tabular}{@{}c|cc@{}}
  & \#UCDs  & \#cEs  \\
	\hline
 \citet{Chilingarian2015}                    & -   & 195 \\  
 \citet{Guerou2015}                          & -   & 7   \\  
 \citet{Ferre-Mateu2018}, (2021)             & -   & 15   \\ 
 \citet{Kim2020}                             & -   & 101 \\  
 \citet{Norris2014}                          & 97 & 39  \\  
 \citet{Liu2015}; \citet{Zhang2018}          & 87  & -   \\  
 \citet{Pandya2016}; \citet{Fahrion2019}     & 364 & -   \\  
 \citet{Dolfi2020}                           & 20  & -   \\  
 \citet{Forbes2020}                          & 9   & -   \\ 
 Other literature                            & 3   & 1   \\  
	\hline
	\hline
\textbf{TOTAL = 937}             & 580 & 357 \\
\end{tabular}
\end{table}  

\begin{table}
\centering
\label{tab:detections}
\footnotesize{}
\caption{Properties of the 30 X-ray detected low-mass compact galaxies.}
\begin{tabular}{lccc}
\hline
\hline 
ID                   & Separation &  $F_\mathrm{0.5-7 keV}$   &  $L_\mathrm{0.5-7 keV}$  \\
                     & (arcsec)   &  (erg s$^{-1}$ cm$^{-2}$) &     (erg s$^{-1}$)       \\
(1)                  & (2)        &        (3)                &           (4)            \\
\hline
FCOS0-2023           &  0.3  &   1.3$^{+0.1}_{-0.2}\times10^{-14}$  &   8.4$^{+0.9}_{-1.0}\times10^{38}$       \\
FCOS1-2095           &  0.1  &   1.3$^{+0.4}_{-0.3}\times10^{-14}$  &   6.1$^{+2.0}_{-1.6}\times10^{38}$       \\
FCOS1-060            &  0.3  &   1.9$^{+2.4}_{-1.2}\times10^{-13}$  &   5.1$^{+6.3}_{-3.2}\times10^{39}$       \\
VHH81-C3             &  0.6  &   9.1$^{+1.1}_{-1.1}\times10^{-15}$  &   7.0$^{+0.8}_{-0.9}\times10^{37}$       \\
Sombrero-UCD1        &  0.5  &   1.0$^{+0.1}_{-0.1}\times10^{-14}$  &   4.2$^{+0.6}_{-0.6}\times10^{38}$       \\
NGC4649\_J67         &  2.8  &   9.2$^{+6.4}_{-3.9}\times10^{-16}$  &   2.8$^{+1.9}_{-1.2}\times10^{37}$       \\
NGC0821-AIMSS1       &  0.7  &   1.5$^{+1.3}_{-1.3}\times10^{-16}$  &   1.1$^{+0.9}_{-0.9}\times10^{37}$       \\
NGC0821-AIMSS2       &  0.5  &   2.5$^{+4.2}_{-1.6}\times10^{-15}$  &   1.9$^{+3.1}_{-1.2}\times10^{38}$       \\
M60-UCD1             &  0.6  &   8.2$^{+7.1}_{-4.0}\times10^{-16}$  &   2.5$^{+2.1}_{-1.2}\times10^{37}$       \\
M87-UCD2             &  1.3  &   3.4$^{+0.3}_{-0.3}\times10^{-15}$  &   1.4$^{+0.1}_{-0.1}\times10^{38}$       \\
M87-UCD6             &  0.9  &   8.0$^{+0.7}_{-0.8}\times10^{-15}$  &   3.2$^{+0.3}_{-0.3}\times10^{38}$       \\
HCH99-18             &  0.1  &   2.1$^{+0.1}_{-0.1}\times10^{-14}$  &   1.5$^{+0.1}_{-0.1}\times10^{38}$       \\
J124352.42+112534.2  &  1.1  &   4.5$^{+0.9}_{-0.9}\times10^{-15}$  &   1.4$^{+0.3}_{-0.3}\times10^{38}$       \\
gregg22              &  1.3  &  9.1$^{+24.1}_{-6.9}\times10^{-15}$  &  4.5$^{+12.0}_{-3.5}\times10^{38}$       \\
gregg31              &  1.9  &   5.4$^{+3.7}_{-3.7}\times10^{-16}$  &   2.7$^{+1.9}_{-1.9}\times10^{37}$       \\
gregg41              &  2.2  &   2.0$^{+0.5}_{-0.5}\times10^{-15}$  &   9.8$^{+2.5}_{-2.5}\times10^{37}$       \\
VCC1499              &  2.3  &   1.1$^{+1.3}_{-0.6}\times10^{-14}$  &   3.2$^{+3.7}_{-1.8}\times10^{38}$       \\
HCH99-16             &  0.2  &   4.4$^{+1.2}_{-1.0}\times10^{-15}$  &   3.2$^{+0.9}_{-0.7}\times10^{37}$       \\
M85-HCC1             &  0.2  &   3.3$^{+0.9}_{-0.9}\times10^{-15}$  &   4.4$^{+1.2}_{-1.2}\times10^{37}$       \\
UCD20                &  0.7  &   2.7$^{+2.3}_{-1.6}\times10^{-15}$  &   2.9$^{+2.5}_{-1.7}\times10^{37}$       \\
HHH86-C15            &  0.8  &   3.2$^{+1.3}_{-1.4}\times10^{-15}$  &   2.3$^{+1.0}_{-1.0}\times10^{37}$       \\
HCH99-21             &  0.9  &   4.7$^{+0.7}_{-0.8}\times10^{-15}$  &   3.5$^{+0.5}_{-0.6}\times10^{37}$       \\
Nuc2\_NGC7727        &  0.4  &   7.3$^{+2.0}_{-1.9}\times10^{-15}$  &   6.3$^{+1.7}_{-1.7}\times10^{38}$       \\
\hline               
NGC4486B             &  0.5  &   4.6$^{+0.9}_{-0.7}\times10^{-15}$  &   2.8$^{+0.5}_{-0.4}\times10^{38}$       \\
PCG012519            &  3.8  &  6.7$^{+10.0}_{-6.7}\times10^{-16}$  &   1.8$^{+2.7}_{-1.8}\times10^{38}$       \\
VCC1178              &  0.2  &   3.0$^{+0.6}_{-0.6}\times10^{-14}$  &   1.1$^{+0.2}_{-0.2}\times10^{39}$       \\
VCC1192              &  0.2  &   1.1$^{+0.2}_{-0.2}\times10^{-14}$  &   5.5$^{+0.9}_{-0.8}\times10^{38}$       \\
M32                  &  0.5  &   2.5$^{+0.2}_{-0.2}\times10^{-14}$  &   1.8$^{+0.1}_{-0.1}\times10^{39}$       \\
NGC~741              &  5.5  &   4.5$^{+0.4}_{-0.4}\times10^{-14}$  &   3.5$^{+0.3}_{-0.3}\times10^{40}$       \\
NGC~2970             &  0.8  &   4.0$^{+1.9}_{-2.0}\times10^{-15}$  &   2.2$^{+1.1}_{-1.1}\times10^{38}$       \\
\hline
\hline
\end{tabular}
\smallskip\newline\small\raggedright {\bf Column designation:}~(1) Galaxy Name; (2) separation between the optical and X-ray position; (3) X-ray flux at 0.5-7 keV; (4) X-ray luminosity at 0.5-7 keV. The solid line separates the UCDs (top) from the cEs (bottom).
\end{table}

\section{Data Analysis}\label{section:data}
\subsection{X-ray emission}
To look for X-ray emission we use the \textit{Chandra} Source Catalog \citep{Evans2010} Version 2 (CSC 2.0) Master Source Catalog. In CSC 2.0 the detection of a source is based on a Maximum Likelihood Estimation
\footnote{\url{https://cxc.cfa.harvard.edu/csc/columns/significance.html}}. We use a rather wide initial search radius of 10 arcsec to account for the range of sizes of the objects (\citealt{Pandya2016} used a more restrictive matching radius of 1.5 arcsec because their sample was limited to UCDs, unlike in this work where cEs are also included). However, we later perform an independent visual inspection of each of the CSC detections to ensure that the position of the X-ray emission falls within the optical extent of the host galaxy in either SDSS, \textit{Hubble Space Telescope}, or DESI Legacy Survey DR8\footnote{\url{https://www.legacysurvey.org/dr8/description/}} images, upon availability. We discard those detections that are clearly not related to the compact galaxy, as is the case of all the UCDs with optical-X-ray position separations above 3 arcsec. 
For those sources with multiple observations, the data of all the $obsids$ in the best Bayesian Block (within which the source is considered to have a constant photon flux\footnote{\url{http://cxc.harvard.edu/csc/data_products/master/blocks3.html}}) are merged in order to measure Point Spread Function (PSF) sizes and model simultaneously the spectral fits. Those sources located in the field or loose groups are fitted by a power-law model with column density $N_\mathrm{H}$ free to vary and with i) photon index $\Gamma$=1.8, typical of AGN and X-ray binaries (e.g. \citealt{Corral2011}; \citealt{Mezcua2018a}), and ii) $\Gamma$ thawed. The model that provides the best statistics (in terms of the reduced statistic \textit{rstat} parameter) within a realistic range of $\Gamma$ when thawed (i.e. $\Gamma$=0-4) is used to compute the spectral flux in the 0.5$-$7 keV band. For those sources located in a cluster we follow the same approach but adding a thermal background model defined by the Astrophysical Plasma Emission Code (APEC; \citealt{Raymond1977}; \citealt{Smith2001}) that describes an emission spectrum from collisionally-ionized diffuse gas with the plasma temperature (kT) free to vary. The models we fit are redshifted models, so that the best fit parameters represent values intrinsic to the source and background. These are then used to get flux corrections to compute rest-frame fluxes. We note that using this approach we are able to derive a spectral flux for sources for which CSC does not provide an aperture flux. The X-ray luminosities in the 0.5-7 keV band are then computed from the spectral flux using the redshift of each source or cluster/group where the source is located (see Table~\ref{tab:detections}).

\subsection{Optical spectroscopy, mid-infrared and radio emission}
We also check whether the sample of low-mass compact galaxies have SDSS spectra available in the DR14 \footnote{\url{https://www.sdss.org/dr14/}}. We note that while it is very unlikely that there will be spectra available for the UCDs, cEs have been found in SDSS in large number (e.g. \citealt{Chilingarian2015}; \citealt{Kim2020}) and the majority have spectra available. This allows us to identify AGN based on emission-line diagnostic diagrams such as the [OIII]/H$_{\beta}$ versus [NII]/H$_{\alpha}$ or [OIII]/H$_{\beta}$ versus [SII]/H$_{\alpha}$ (i.e. BPT diagram; \citealt{Baldwin1981}; \citealt{Kewley2001,Kewley2006}; \citealt{Kauffmann2003}). To apply these diagnostics, we consider only those sources whose H$_{\alpha}$, H$_{\beta}$, [NII]$\lambda$6583, [SII]$\lambda$6717,6731, and [OIII]$\lambda$5007 emission lines have a signal-to-noise ratio $\geq$3. The remaining sources are considered as quiescent (e.g. \citealt{Mezcua2020}). 

\begin{figure*}
\centering
\includegraphics[width=\textwidth]{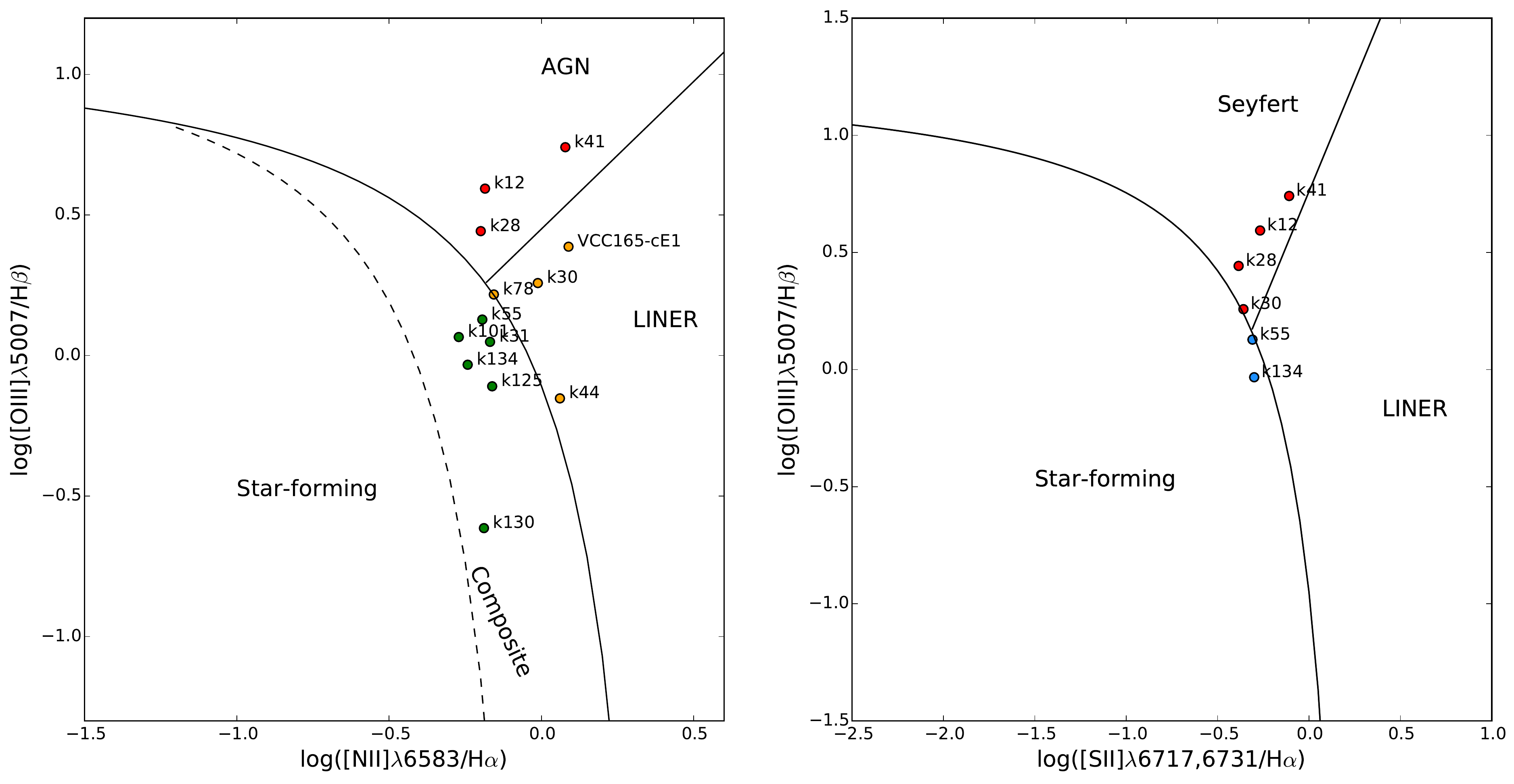}
\caption{[OIII]/H$_{\beta}$ versus [NII]/H$_{\alpha}$ (left) and [OIII]/H$_{\beta}$ versus [SII]/H$_{\alpha}$ (right) emission line diagnostic diagram for those low-mass compact galaxies with signal-to-noise ratio $\geq$3 in the emission lines used in these diagrams.}
\label{BPT}
\end{figure*}

To identify AGNs in the mid-infrared regime we search for WISE counterparts at 3.6$\mu$m, 4.5$\mu$m and 5.8$\mu$m or all the low-mass compact galaxies in our sample and apply the mid-infrared color-color cut of \cite{Jarrett2011} and \cite{Stern2012}. We also search for radio emission at 1.4 GHz in the FIRST\footnote{Faint Images of the Radio Sky at Twenty Centimeters.} \citep{Becker1995} survey using a search radius of 5 arcsec, which is the angular resolution of the survey. In all cases, as done for the X-ray detections, those objects with detections have been visually inspected to confirm that the signature belongs to the compact source and not neighbouring or background objects.

\section{Results}\label{section:results}
Out of the 937 low-mass compact galaxies compiled in this paper, 566 have \textit{Chandra} observations (90 cEs and 476 UCDs). Of these, 30 are detected by the CSC 2.0, being 7 of them X-ray detections in cEs and 23 in UCDs. Before discussing the detection rates, we must caution about the selection bias that is inherent to our sample. The majority of UCDs in our sample belong to clusters because these are the places where most works have looked for them. Nearly 70\% (16 out of 23) of the X-ray detected UCDs are in dense environments, which could explain the possible presence of gas available for accretion. Instead, the majority of the cEs belong to SDSS compilations and thus are more biased towards the field environment. Nearly 86\% (6 out of 7) of the X-ray detected cEs are in the field or belong to loose groups, where they should be less prone to have gas available for accretion. 

Nevertheless, the X-ray detection rates amongst UCDs and cEs are remarkably similar, being 7/90=7.8\% for the cEs and of 23/476=4.8\% for the UCDs assuming a detection limit $L_\mathrm{0.5-7 keV} > 2 \times 10^{37}$ erg s$^{-1}$ (the lowest value detectable in our sample). A detection rate of $\sim$3$\%$ was found for the UCDs in \citet{Pandya2016}, which is lower than our value. \cite{Hou2016} also report a detection rate of $\sim$3$\%$, but they include both UCDs and extended stellar clusters. Considering the same conditions as in \citeauthor{Pandya2016} (completeness limit of $L_\mathrm{0.5-7 keV} > 2 \times 10^{38}$ erg s$^{-1}$ and optical-to-X-ray position separations within 1.5 arcsec) we would obtain an X-ray detection rate of 1.5\% for UCDs, closer to that of \cite{Pandya2016} but slightly lower given our larger number of UCDs even when applying their conditions (461 sources compared to the 149 of \citealt{Pandya2016}). 

\subsection{Origin of the X-ray emission}
To search for AGNs amongst the 30 X-ray detected cEs and UCDs of Table~\ref{tab:detections}, we apply a luminosity threshold of $L_\mathrm{0.5-7 keV} \geq 10^{40}$ erg s$^{-1}$, of the same order as the X-ray luminosity of the faintest AGN in dwarf galaxies (e.g. \citealt{Baldassare2017}; \citealt{Mezcua2018a}) and of those ultraluminous X-ray sources hosting IMBHs (e.g. \citealt{Kim2015}; \citealt{Mezcua2018b}). This results in one source: the cE NGC\,741, which is a well-studied AGN at X-ray, optical, and radio wavelengths (e.g. \citealt{Schellenberger2017}; \citealt{Kim2019}) with a BH mass upper limit of log\mbh $<$ 8.67 derived from gas dynamics using \textit{Hubble Space Telescope}/STIS spectroscopy (\citealt{Beifiori2009}; \citealt{vandenBosch2016}). 
 
The remaining sources have X-ray luminosities in the range $L_\mathrm{0.5-7 keV} \sim 10^{37} - 5 \times 10^{39}$ erg s$^{-1}$, and are thus consistent with being either X-ray binaries or ultraluminous X-ray sources powered by stellar-mass BHs or neutron stars with super-Eddington accretion (e.g. \citealt{Bachetti2014}; \citealt{Fuerst2016}; \citealt{Israel2017}). This would be in agreement with the previous findings that the X-ray emission of UCDs is consistent with that from (low-mass) X-ray binaries (e.g. \citealt{Pandya2016}; \citealt{Hou2016}). We caution, however, that a weak level of SMBH accretion cannot be fully discarded at such X-ray luminosities. Assuming that low-mass compact galaxies with $L_\mathrm{0.5-7 keV} \sim 10^{37} - 5\times10^{39}$ erg s$^{-1}$ could be SMBHs accreting at an Eddington ratio of $\lambda=10^{-6.5}$, as confirmed for M60-UCD1 (\citealt{Hou2016}), we can use the relation \mbh = ($L_\mathrm{bol}\lambda)/1.3\times10^{38}$, where $L_\mathrm{bol}=k\times L_\mathrm{0.5-7 keV}$, assuming a bolometric correction factor of $k$ = 10 (e.g. \citealt{Duras2020}). We caution the reader about using these estimates as they rely on some assumptions that cannot be tested here. Comparing the estimated values with those for which there is a literature measurement, we find that while many are in fair agreement, others show almost two orders of magnitude difference, being our estimate typically larger than the measured. This means that these values should be treated, at most, as an upper limit. With this in mind, we find that the low-mass compact galaxies have BH masses in the range log \mbh = 6.5--9.0 \msun, this is, more compatible with being a SMBH rather than an IMBH. This would support the stripped origin for the UCDs but is less clear for the seven cEs, which have BH estimates of $7.5$--$8.5$\msun. We will further discuss these assumptions in Section~\ref{section:discussion}.

\begin{table}
\centering
\label{tab:AGNtab}
\footnotesize{}
\caption{Properties of the eleven cEs hosting an AGN.}
\begin{tabular}{lcccccccc}
\hline
\hline 
ID                   &  SDSS  & $L_\mathrm{0.5-7 keV}$             & $L_\mathrm{1.4 GHz}$   \\
                     &        &  (erg s$^{-1}$)                    &   (erg s$^{-1}$)       \\
(1)                  &  (2)   &     3)                             &          (4)           \\
\hline               
NGC~741$\dagger$              &  --    &   3.5$^{+0.3}_{-0.3}\times10^{40}$ & 1.4$\times10^{38}$     \\
J085431.18+173730.5$\dagger\dagger$  &  QSO   &        --                          &           --           \\
J122427+271407       &  --    &        --                          &   1.1$\times10^{39}$   \\
k12                  &  AGN   &        --                          &       --               \\
k28                  &  AGN   &        --                          &       --               \\
k30                  &  AGN   &        --                          &       --               \\
k41                  &  AGN   &        --                          &       --               \\
k55                 &  AGN   &        --                          &       --               \\
k123                 &   --   &        --                          &   1.7$\times10^{38}$   \\
k134                 &  AGN   &        --                          &       --               \\
VCC165-cE1           &  AGN   &        --                          &       --               \\
\hline
\hline
\end{tabular}
\smallskip\newline\small\raggedright {\bf Column designation:}~(1) Galaxy Name; (2) optical classification of the galaxy based on SDSS spectroscopy; (3) \textit{Chandra} X-ray luminosity at 0.5-7 keV; 
(4) FIRST radio luminosity at 1.4 GHz.\\
$\dagger$ For a detailed X-ray and radio analysis see \cite{Schellenberger2017}.\\ 
$\dagger\dagger$ See \cite{Paudel2016}.
\end{table}

\begin{figure}
\centering
\includegraphics[width=0.48\textwidth]{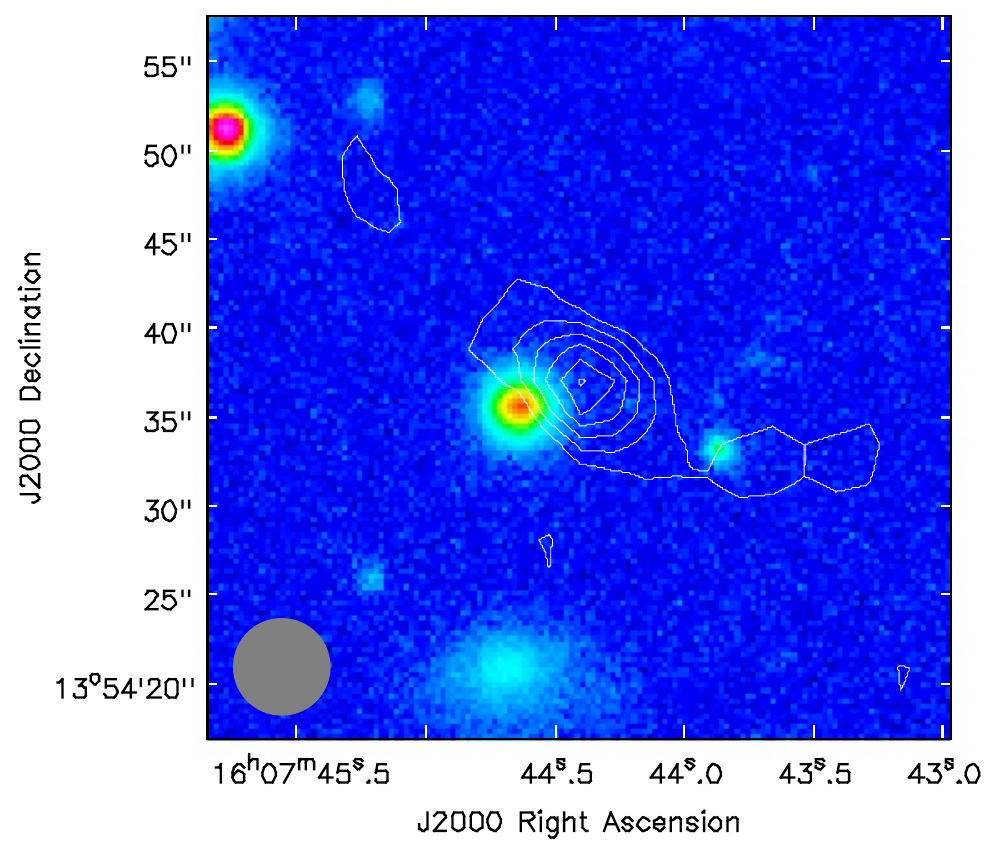}
\caption{DESI Legacy Survey DR8 image of k123 with the FIRST radio contours overplotted in white. Contours start at 3 times the root-mean-square of 0.149 mJy beam$^{-1}$. The restoring beam of 5.4 arcsec radius is shown as a grey circle.}
\label{FIRST}
\end{figure}

\subsection{Other wavelengths}
Of the 937 low-mass compact galaxies 312 have SDSS data, with only three of them being UCDs. This bias towards cEs is due to cEs being more massive and brighter than UCDs. Therefore, in most cases there are SDSS images available for them but not spectroscopy. Additionally, UCDs are found in large numbers inside clusters, so there is also the inherent bias of SDSS towards observations in the field. The different emission-line diagnostic diagrams are shown in Figure~\ref{BPT}. Of the 312 low-mass compact galaxies with SDSS spectra, 13 have emission lines that would fulfill the signal-to-noise criteria required for the [OIII]/H$_{\beta}$ \textit{vs.} [NII]/H$_{\alpha}$ diagram (left panel). 6 of them also fulfill the signal-to-noise criteria required for the [OIII]/H$_{\beta}$ \textit{vs.} [SII]/H$_{\alpha}$ diagram (right panel), and one (J085431.18+173730.5) is a known quasar with broad emission lines \citep{Paudel2016}. 

Three sources qualify as AGN in the [OIII]/H$_{\beta}$ versus [NII]/H$_{\alpha}$, all of them being also an AGN in the [OIII]/H$_{\beta}$ versus [SII]/H$_{\alpha}$ (k12, k28, k41; shown in red in Figure~\ref{BPT}). Three additional sources are classified as LINERs\footnote{Low Ionisation Emission Line Region.} in the [OIII]/H$_{\beta}$ versus [NII]/H$_{\alpha}$
(k30, k44, VCC165-cE1), one of which is an AGN in the [OIII]/H$_{\beta}$ versus [SII]/H$_{\alpha}$. Because the LINER emission can be produced not only by AGN but also by hot old stars with typically an H$_{\alpha}$ equivalent width EW(H$\alpha$) $<$ 3\AA, we apply a diagnostic diagram EW(H$\alpha$) vs [NII]/H$_{\alpha}$ (i.e. the WHAN diagram; \citealt{CidFernandes2010}) to distinguish between the two processes. According to the WHAN diagram, two of the three LINERs qualify as AGN (k30 and VCC165-cE1). We also check how many of the Composite objects in the [OIII]/H$_{\beta}$ versus [NII]/H$_{\alpha}$ BPT, this is, those whose emission lines are ionized by a combination of both AGN and star formation processes, are classified as AGN in the WHAN diagram. We find that two of the Composites (k55 and k134) qualify as AGN according to the WHAN diagram.

The three BPT AGN and the four WHAN AGN are all cEs and have been also identified as AGN by Rey, Oh, \& Kim (2021, submitted) and \cite{Bar2017}. We refer the reader to these publications for a more detailed study of these AGN. They have not been observed by \textit{Chandra}. 

The total number of SDSS AGN candidates from our sample is thus of eight (seven based on the BPT and WHAN diagrams, one classified as a QSO with broad emission lines; \citealt{Paudel2016}), being all of them cEs (marked as AGN or QSO in Table~3). For J085431.18+173730.5, \cite{Paudel2016} derive a BH mass of 2.1$\times10^{6}$ \msun from the luminosity and width of the broad H$\alpha$ emission line in the SDSS spectrum. For the remaining sources, we can estimate the BH mass using the scaling relation between the narrow $L_\mathrm{[OIII]}/L_\mathrm{H\beta}$ line ratio and the width of the broad H$\alpha$ emission line of \cite{Baron2019}, which allows estimating BH masses from narrow emission lines only and has an uncertainty of at least 0.5 dex. Combining eqs. 1, 5, and 6 of \cite{Baron2019} we find:
\begin{equation}
\label{eq1}
log M_{BH} = log \epsilon + 3.55log(L_{[OIII]}/L_{H\beta}) + 0.59log L_{bol}-20.96  
\end{equation}
where we use a scale factor $\epsilon$=1.075 as in \cite{Baron2019}. The bolometric luminosity is derived as (\citealt{Netzer2009}):
\begin{equation}
log L_{bol} = log L_{H\beta} +3.48 + max[0,0.31(log [OIII]/H\beta -0.6)]
\end{equation}
where $L_\mathrm{H\beta}$ has been extinction-corrected using the reddening curves of \cite{Calzetti2000}. 
Using eq.~\ref{eq1} we find a range of BH masses for the SDSS AGN candidates of log $M_\mathrm{BH}$ = 4.6-7.0 M$_{\odot}$. Note that these values are on average lower than the ones obtained from the $L_\mathrm{0.5-7 keV}$ for the seven X-ray detected cEs in Section~\ref{section:results}. Unfortunately, none of the cEs classified as AGN based on SDSS spectroscopy has an X-ray detection to compare the results. As with the BH mass estimated from the X-ray luminosity above, we caution the reader about using the values derived here in Eq.~\ref{eq1} and should only be considered as first order approximations.

In the mid-infrared regime, 38 out of the 937 galaxies have a WISE counterpart; however, none of them qualifies as an AGN according to the color-color cuts of \cite{Jarrett2011} and \cite{Stern2012}. Three out of the 937 galaxies have a FIRST counterpart within $\sim$5 arcsec: the cEs NGC\,741-AIMSS1, k123 and J122427+271407. The radio emission is resolved in the three sources, with NGC\,741-AIMSS1 and k123 showing extended, jet-like radio structures suggestive of AGN radio emission (see Fig.~\ref{FIRST}). NGC\,741-AIMSS1 is indeed a known radio galaxy (e.g. \citealt{Schellenberger2017}). k123 and J122427+271407 have integrated fluxes at 1.4 GHz of 4.43 mJy and 2.34 mJy, respectively, which correspond to radio luminosities of $L_\mathrm{1.4 GHz}=1.7\times10^{38}$ erg s$^{-1}$ and $L_\mathrm{1.4 GHz}=1.1\times10^{39}$ erg s$^{-1}$, respectively, at the distance of the galaxies (see Table~3). They have a negligible star-formation rate from SDSS spectra. We note that the radio luminosities are of the same order as that of radio AGN in dwarf galaxies after removing the contribution from star formation to the radio emission ($L^\mathrm{AGN}_\mathrm{1.4 GHz}\sim10^{38-40}$ erg s$^{-1}$; e.g. \citealt{Mezcua2019}; \citealt{Reines2020}). Based on all the above, we thus consider k123 and J122427+271407 as strong AGN candidates. Our sample of possible AGN candidates thus contains a total of eleven objects (Table~3).

\begin{table*}
\centering
\label{tab:final_sample}
\footnotesize{}
\caption{Compilation of low-mass compact galaxies with BH mass measurements or estimates}
\begin{tabular}{lccccccccccc}
\hline
\hline 
Galaxy               &      RA       &   DEC        & X-ray  & AGN    &  Re     &  $\sigma$ &  \Mstar  &  \mbh$_{Lit}$ & \mbh$_{L_\mathrm{X}}$ & \mbh$_{Eq.1}$   & \mbh$_{G20}$   \\             
                     &  (J2000)      & (J2000)      & det.   & SDSS   &  (pc)   &    (\kms) &  (\msun) &   (\msun)     &  (\msun)                      &  (\msun)        &  (\msun)        \\ 
(1)                  &    (2)        &    (3)       & (4)    &  (5)   &  (6)    &     (7)   &    (8)   &   (9)         &  (10$\S$)                &   (11$\S$) &   (12$\S$)  \\
\hline                                                                                                                                  
FCOS0-2023           &  03:38:12.70  &  -35:28:57.0 &   Y    &  N     &   4.0   &   15.3    &   6.7    &    --         &  8.3                          &  --            &   3.9    \\             
FCOS1-2095           &  03:38:33.82  &  -35:25:57.0 &   Y    &  N     &   3.3   &   28.0    &   6.7    &    --         &  8.2                          &  --            &   5.1    \\             
FCOS1-060            &  03:39:17.67  &  -35:25:30.2 &   Y    &  N     &  10.0   &   29.1    &   7.0    &    --         &  9.1                          &  --            &   5.2    \\             
VHH81-C3             &  13:24:58.21  &  -42:56:10.0 &   Y    &  N     &   4.2   &   16.1    &   6.5    &    --         &  7.2                          &  --            &   4.1    \\             
Sombrero-UCD1        &  12:40:03.13  &  -11:40:04.3 &   Y    &  N     &  14.7   &   31.9    &   7.2    &    --         &  8.0                          &  --            &   5.4    \\             
NGC4649\_J67         &  12:43:38.50  &  +11:33:02.7 &   Y    &  N     &   9.8   &    --     &   6.2    &    --         &  6.8                          &  --            &   --     \\             
NGC0821-AIMSS1       &  02:08:20.70  &  +10:59:26.6 &   Y    &  N     &   6.5   &    --     &   6.7    &    --         &   --                          &  --            &   --     \\             
NGC0821-AIMSS2       &  02:08:20.70  &  +10:58:55.5 &   Y    &  N     &   8.4   &    --     &   6.9    &    --         &  6.4                          &  --            &   --     \\             
M60-UCD1             &  02:43:36.00  &  +11:32:04.6 &   Y    &  N     &  27.2   &   71.0    &   8.3    &   7.3         &   --                          &  --            &   6.9    \\             
M87-UCD2             &  02:30:48.00  &  +12:24:33.0 &   Y    &  N     &   --    &    --     &    --    &    --         &  7.5                          &  --            &   --     \\             
M87-UCD6             &  02:30:47.54  &  +12:24:24.1 &   Y    &  N     &   --    &    --     &    --    &    --         &  7.9                          &  --            &   --     \\             
HCH99-18             &  03:25:31.60  &  -43:00:02.8 &   Y    &  N     &   --    &    --     &    --    &    --         &  7.6                          &  --            &   --     \\             
J124352.42+112534.2  &  12:43:52.42  &  +11:25:34.2 &   Y    &  N     &   --    &    --     &    --    &    --         &  7.5                          &  --            &   --     \\             
gregg22              &  03:38:09.27  &  -35:35:07.0 &   Y    &  N     &   --    &    --     &    --    &    --         &  8.0                          &  --            &   --     \\             
gregg31              &  03:38:21.84  &  -35:25:13.8 &   Y    &  N     &   --    &    --     &    --    &    --         &  6.8                          &  --            &   --     \\             
gregg41              &  03:38:36.99  &  -35:25:44.2 &   Y    &  N     &   --    &    --     &    --    &    --         &  7.4                          &  --            &   --     \\             
VCC1499              &  12:33:19.80  &  +12:51:12.0 &   Y    &  N     &   --    &    --     &    --    &    --         &  7.9                          &  --            &   --     \\             
HCH99-16             &  13:25:30.29  &  -42:59:34.8 &   Y    &  N     &   6.5   &    --     &   6.7    &    --         &  6.9                          &  --            &   --     \\             
M85-HCC1             &  12:25:22.84  &  +18:10:53.7 &   Y    &  N     &   --    &    --     &    --    &    --         &  7.0                          &  --            &   --     \\             
UCD20                &  10:05:17.02  &  -07:43:52.7 &   Y    &  N     &   --    &    --     &    --    &    --         &  6.8                          &  --            &   --     \\             
HHH86-C15            &  13:25:30.41  &  -43:11:49.6 &   Y    &  N     &   --    &    --     &    --    &    --         &  6.8                          &  --            &   --     \\             
HCH99-21             &  13:25:34.65  &  -43:03:27.7 &   Y    &  N     &   --    &    --     &    --    &    --         &  6.9                          &  --            &   --     \\             
Nuc2\_NGC7727        &  23:39:53.80  &  -12:17:34.0 &   N    &  N     &   27.4  &    79.0   &   8.2    &   6.5         &  8.2                          &  --            &   7.1    \\             
FornaxUCD3           &  03:38:54.0   &  -35:33:34.0 &   N    &  N     &   87.2  &    25.0   &    --    &   6.1         &   --                          &  --            &   4.9    \\             
S999                 &  12:30:45.91	 & +12:25:01.8  &   N    &  N     &   21.3  &    26.0   &   6.6    &   7.4$\dagger$&  8.2                          &  --            &   5.0    \\             
VUCD3                &  12:30:57.40	 & +12:25:44.8  &   N    &  N     &   18.6  &    33.0   &   7.6    &   6.6         &  6.8                          &  --            &   5.4    \\             
M59cO                &  12:41:55.33	 & +11:40:03.7  &   N    &  N     &   32.1  &    29.0   &   8.3    &   6.8         &   --                          &  --            &   5.2    \\             
M59UCD3              &  12:42:11.05  &  +11:38:41.3 &   N    &  N     &   25.0  &    78.0   &   8.6    &   6.6         &   --                          &  --            &   7.1    \\             
UCD320               &  13:25:52.70  &  -43:05:46.6 &   N    &  N     &    6.8  &    19.0   &   6.4    &   6.0$\dagger$&   --                          &  --            &   4.4    \\             
UCD330               &  13:25:54.30  &  -43:05:46.6 &   N    &  N     &    3.3  &    30.5   &   6.7    &   5.0$\dagger$&   --                          &  --            &   5.3    \\             
NGC4546-UCD1         &  12:35:28.70	 & -03:47:21.1  &   N    &  N     &   25.5  &    21.8   &   7.6    &   6.0$\dagger$&   --                          &  --            &   4.6    \\             
\hline                                                                                                                                                
NGC4486B             &  12:30:31.97  &  +12:29:24.6 &   Y    &   N    &  180.0  &   170.0   &   9.5    &   7.9         &  7.8                          &   --           &   8.5    \\             
PCG012519            &  03:20:32.90  &  +41:34:26.8 &   Y    &   N    &  390.0  &   222.1   &   10.    &    --         &  7.6                          &   --           &   9.0    \\             
VCC1178              &  12:29:21.29  &  +08:09:23.8 &   Y    &   N    &  486.0  &   125.0   &   9.9    &   7.4         &  8.4                          &   --           &   7.9    \\             
VCC1192              &  12:29:30.25  &  +07:59:34.3 &   Y    &   N    &  409.0  &    16.5   &   9.1    &   6.0         &  8.1                          &   --           &   4.1    \\             
M32                  &  00:42:41.80  &  +40:51:55.0 &   Y    &   N    &  113.0  &    76.0   &   9.0    &   6.5         &  8.6                          &   --           &   7.0    \\             
NGC~741              &  01:56:21.30  &  +05:37:46.8 &   Y    &   Y    &  311.7  &    86.2   &   9.8    &   8.7         &  9.9                          &   --           &   7.2    \\             
NGC~2970             &  09:43:31.10  &  +31:58:37.1 &   Y    &   N    &  793.0  &    47.4   &   9.1    &    --         &  7.7                          &   --           &   6.1    \\             
J085431.18+173730.5  &  08:54:31.18  &  +17:37:30.5 &   N    &   Y    &  490.0  &     --    &   9.2    &   6.3         &   --                          &   --           &   --     \\             
J122427+271407       &  12:24:27.57  &  +27:14:07.5 &   N    &   Y    &  --     &    --     &   --     &    --         &   --                          &   --           &   --     \\             
k12                  &  03:43:30.26  &  -07:35:07.4 &   N    &   Y    &  381.2  &    89.1   &   9.2    &    --         &  8.6                          &  6.3           &   7.3    \\                 
k28                  &  09:37:14.74  &  +37:17:30.5 &   N    &   Y    &  255.3  &    96.7   &   9.2    &    --         &  8.7                          &  6.1           &   7.5    \\                   
k30                  &  09:44:51.02  &  +12:30:44.3 &   N    &   Y    &  397.8  &    77.7   &   9.0    &    --         &  8.8                          &  5.3           &   7.0    \\                     
k41                  &  10:39:18.55  &  +17:11:17.2 &   N    &   Y    &  385.3  &    75.7   &   9.2    &    --         &  8.9                          &  7.0           &   6.9    \\                              
K55                  &  11:27:30.82  &  +03:09:58.0 &   N    &   Y    &  582.0  &    67.5   &   9.3    &    --         &   --                          &  4.8           &   6.8    \\
k123                 &  16:07:44.64  &  +13:54:35.6 &   N    &   Y    &  281.1  &   102.9   &   8.9    &    --         &   --                          &   --           &   7.6    \\             
k134                 &  22:59:58.34  &  +17:55:16.0 &   N    &   Y    &  528.0  &   167.7   &   9.5    &    --         &  8.9                          &  4.6           &   8.5    \\             
VCC165-cE1           &  12:15:51.26  &  +13:13:03.4 &   N    &   Y    &  200.0  &    88.8   &   9.6    &    --         &  8.6                          &  5.5           &   7.3    \\                
\hline                                                                                                                                                                       
\hline
\end{tabular}
\smallskip\newline\small\raggedright {\bf Column designation:}~(1) Galaxy Name; (2,3) RA, DEC coordinates of the optical galaxy; (4,5) If the object has been detected in the X-ray section; (5) if the object is an AGN candidate from the indicators in Section~4; (6) Literature effective radii; (7) Literature velocity dispersion; (8) Literature stellar mass; (9) Literature BH mass ($\dagger$ should be considered as upper limits); (10) BH mass estimated from Eq. 1(this work); (11) BH estimated from the \citealt{Baron2019} relation (this work); (12) BH estimated from the \mbh-$\sigma$ relation of G20. The solid line separates the UCDs (top) from the cEs (bottom).)\\
$\S$ \textit{We highly caution the reader to not use them as proper values but estimates}.
\end{table*}

\subsection{AGN fraction}
Amongst the 937 low-mass compact galaxies analyzed in this paper, 11 out of 357 cEs (but none of the UCDs) qualify as AGN from different indicators (see Table~3). Albeit of completeness effects, the AGN fraction amongst low-mass compact galaxies is therefore 1.2\% (11/937). This is significantly lower than that of red massive galaxies at low redshifts ($\sim$10\%; e.g. \citealt{Gu2018}) and than that of galaxies hosting both NSCs and AGN ($>$10\%; e.g. \citealt{Seth2008}), as expected given the higher stellar masses covered by these studies (log M$_{*} > 9$ M$_{\odot}$). When considering only the X-ray selected AGN, the AGN fraction in low-mass compact galaxies is 0.1\% (1/937), which is of the same order as that of X-ray selected AGN in dwarf galaxies at z $\lesssim$ 0.3 ($\sim$0.4\%; \citealt{Mezcua2018a}; \citealt{Birchall2020}). Adding the weakly accreting SMBHs already published for the cE M32 (\citealt{Peng2020}) and the UCDs M60-UCD1, M59-UCD3, and NGC7727-Nuc 2 (\citealt{Schweizer2018}), the AGN fraction amongst compact galaxies rises to 1.6\% (15/937). 

\begin{figure*}
\centering
\includegraphics[width=1.02\textwidth]{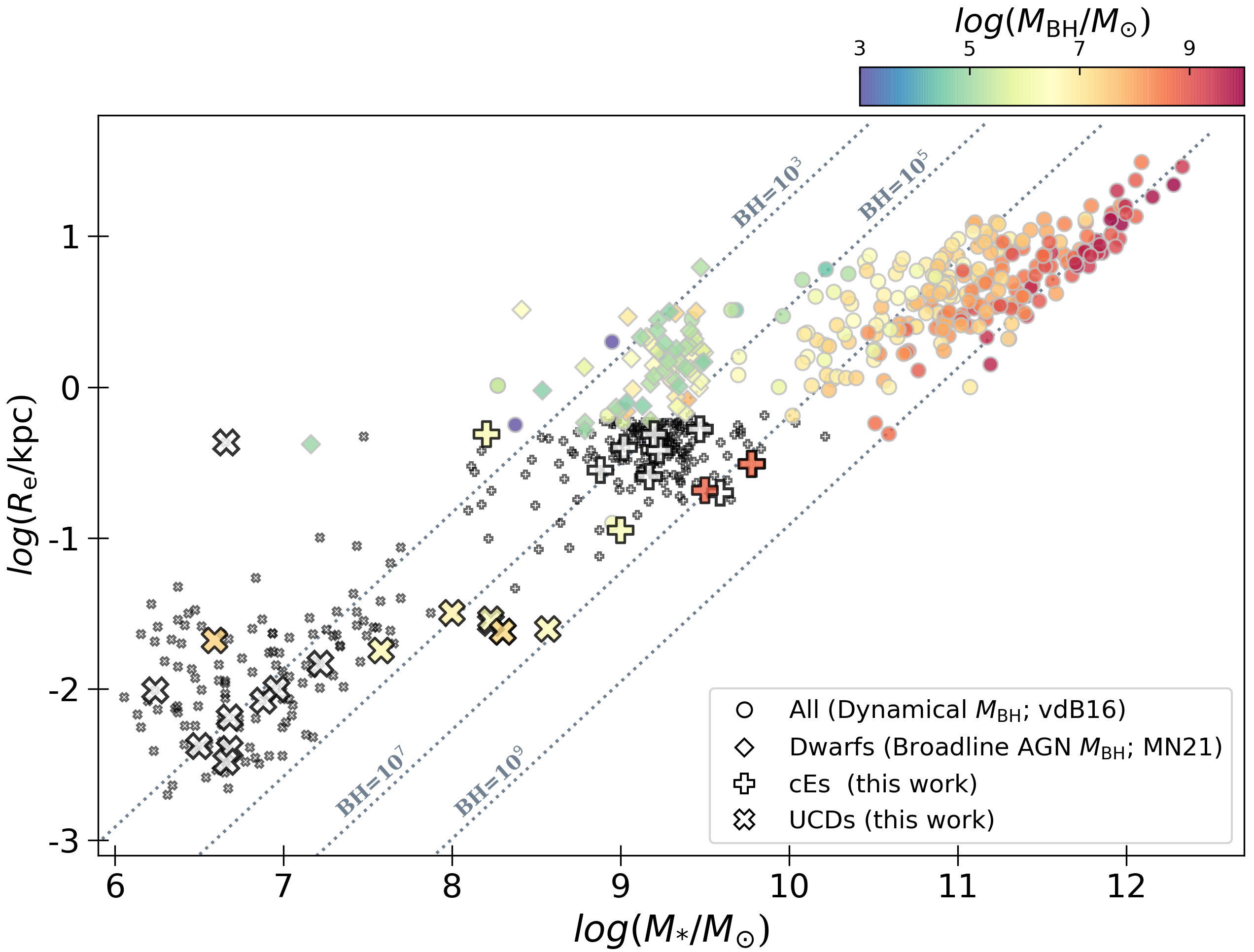}
\caption{Mass-size relation for galaxies at different stellar masses, also showing the fundamental plane of \Mstar-$\re$-\mbh with lines of iso-\mbh\, (dotted; increasing from left to right). The figure shows galaxies of different types as control sample: ETGs and dwarfs from vdB16 (filled circles), and dwarf galaxies from MN21 (filled diamonds). Our sample of low-mass compact galaxies is represented by small crosses (cEs) and small exes (UCDs), when masses and sizes are available. Those galaxies with an X-ray detection or that are AGN candidates (Tables~2 and 3) or that have a BH mass measurement reported in the literature are shown with a larger symbol. For all galaxy types, the color scale of the symbols corresponds to the direct measurements of their BH masses. While cEs roughly follow the lines of iso-\mbh, the BH masses of the UCDs are larger than expected from the relations, which would suggest a stripping origin for these galaxies. However, they appear $\sim$3-4 orders of magnitude higher than expected, suggesting that some other effect might be needed to explain such large BH masses}.
\label{figure:mass_size}
\end{figure*}

\section{Black holes in low-mass compact galaxies: super-massive or intermediate-mass?}
\label{section:discussion}
ETGs are known to follow several scaling relations. One of these is the mass-size relation, which suggests that more massive ETGs tend to be larger than low-mass ones. This is tightly followed at the high-mass end in particular (e.g.  \citealt{Gallazzi2005,Gallazzi2020}), but it branches out into two different regimes at a characteristic stellar mass scale of $\sim3\times10^{10}$\msun\, (e.g. \citealt{Baldry2012}; \citealt{Cappellari2016}). The branch corresponding to the more extended low-mass galaxies (dwarf ellipticals and dwarf spheroidals) flattens from that point down, while the branch of low-mass compact galaxies covering cEs, UCDs and down to globular clusters seems to follow the trend of the massive ETGs downwards (albeit showing a larger scatter; e.g. \citealt{Drinkwater2000}; \citealt{Brodie2011}). 

Furthermore, ETGs follow rather tight relations for some of the main galaxy properties (like stellar mass, stellar velocity dispersion, bulge luminosity or NSC mass) with their BH mass (e.g. \citealt{Kormendy2013}). Because all these properties are correlated, a Fundamental Plane for the BH scaling relations was defined in \hypertarget{vdB16}{\citet{vandenBosch2016}} (\hyperlink{vdB16}{vdB16} hereafter; eq. 5), correlating the BH mass with the mass--size relation mentioned above. However, the behaviour of this fundamental plane at the low-mass end is a matter of high debate and unfortunately, the number of low-mass compact objects with measured BH masses, necessary to tackle this issue, is rather scarce. 

To circumvent this, we have compiled a sample of low-mass compact object with measured and/or estimated BH masses by using several scaling relations and methods. They are presented in Table~4, along with their most relevant properties for this study: their sizes, stellar masses, and BH masses from both direct measurements reported in the literature and the new estimates performed here (although, as mentioned before, we caution the reader about using these estimates as BH measurements). The table contains all the low-mass compact galaxies we have found to have an X-ray detection (Sect. 4.1), those compatible with being an AGN (Sect. 4.2) and also includes other low-mass compact galaxies that have direct measurements of their BHs. This gives a total of 31 UCDs and 16 cEs that will be used in this section.

Figure \ref{figure:mass_size} shows the \Mstar-$\re$-\mbh fundamental plane, marked by the dotted lines, increasing from left to right from \mbh$\sim10^{3}$\msun\, to $\sim10^{9}$\msun. Overplotted with filled dots we show a sample of ETGs, mostly comprising massive ellipticals and some dwarfs ellipticals. They are color-coded by the BH mass measurement published in \hyperlink{vdB16}{vdB16}. These are direct measurements, not estimates, but it is clear that they closely follow the lines marked by the \mbh\, fundamental plane. Dwarf galaxies with BH masses measured from AGN broad emission lines (from Mart\'in-Navarro et al. submitted; MN21 hereafter) are also included and shown by filled diamonds. Small symbols (crosses for the cEs, exes for the UCDs) correspond to the compilation presented in Section~\ref{section:sample}, when size and mass are available. The larger symbols of crosses and exes correspond to the final sample of low-mass compact galaxies from Table~4, selected because they are an X-ray detection, AGN candidates, or objects for which a BH mass has already been measured. For the latter, they are color-coded accordingly following the color-bar. 

If low-mass compact galaxies were to be formed intrinsically (and thus follow the scaling relations at the low-mass end), the expectations are that UCDs should roughly follow the \mbh $\sim10^{3-4}$\msun\, dotted lines and cEs the $\sim10^{5-7}$\msun\, ones. However, it is clear that low-mass compact galaxies overall present BH masses that are ubiquitously larger than expected from such relations. See, for example, the compact galaxies around the \mbh$\sim$10$^{3}$\msun\, line, which can show up to 4 orders of magnitude difference between the measured and the expected BH mass. We note the caveat that most of the measured BH masses have been found so far in high-mass UCDs (\Mstar$\gtrsim 10^{7}$\msun), with values of BH mass more compatible with a SMBH than an IMBH. Overall, cEs seem to be less extreme, being roughly one order of magnitude larger, which is still compatible with a stripped origin. In fact, some of them follow closely the lines of iso-\mbh\,, which would suggest an intrinsic origin for these cases. This supports what has been seen from the study of their stellar populations and kinematic: cEs are mixed-bag of both origins (e.g. \citealt{Ferre-Mateu2018}; \citealt{Kim2020}; \citealt{Ferre-Mateu2021}).

\begin{figure}
\centering
\includegraphics[scale=0.4]{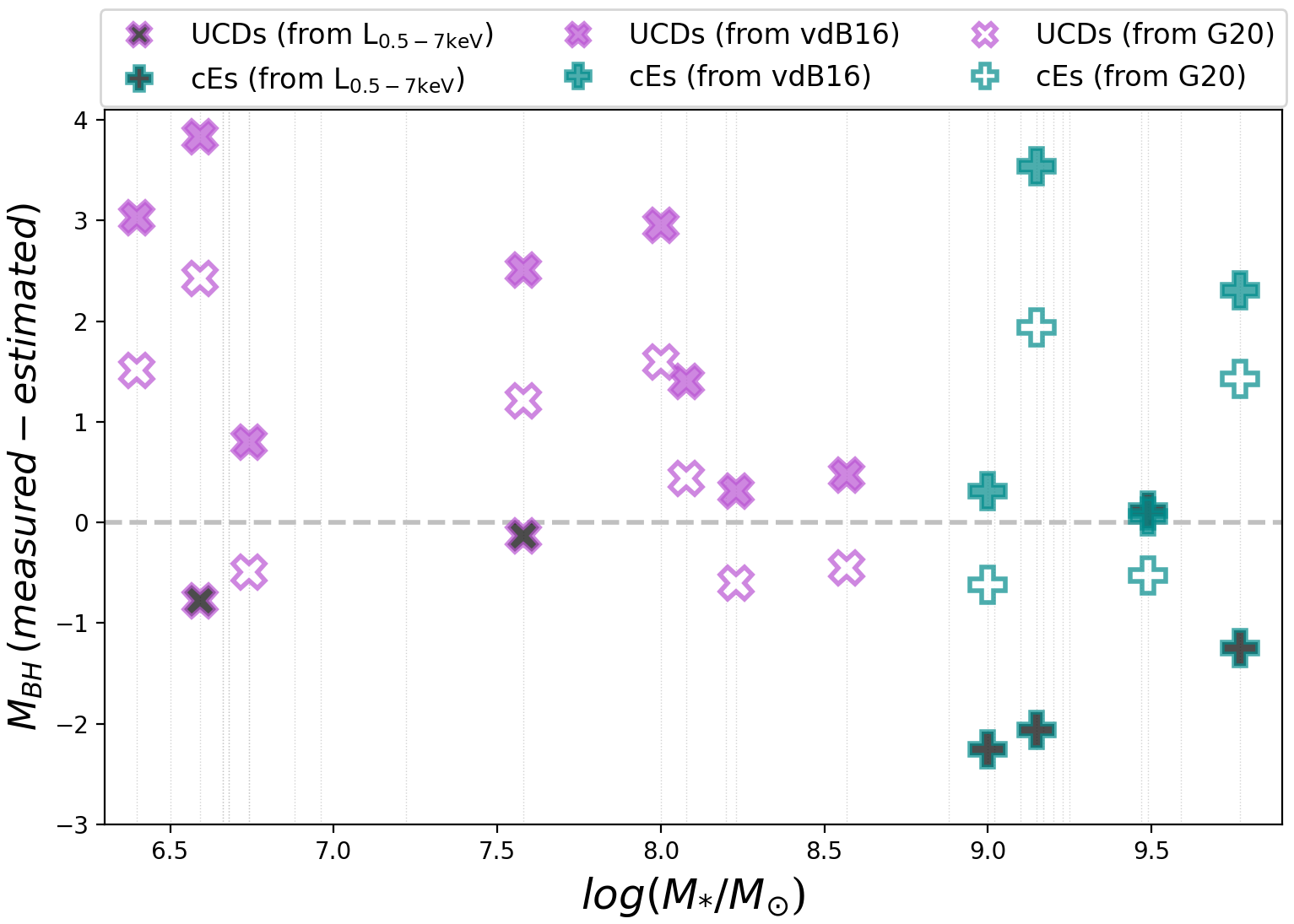}
\caption{Differences between the direct measurement and estimated BH masses based on different relations (for those low-mass compact galaxies with direct measurements available). Black crosses with colored outline correspond to the BH mass estimates from the X-ray luminosities (Sect.~4.1). Filled and open colored symbols correspond to the \mbh-$\sigma$ relations of vdB16 and G20, respectively, to estimate the BH mass.} 
\label{figure:difference_BHs}
\end{figure}

The differences between the measured an the estimated BH masses is better seen in Figure~\ref{figure:difference_BHs}. It shows, at a given galaxy's stellar mass, the difference between the direct measurement and the  estimate for the BH mass. We consider the BH mass estimates for the galaxies with an X-ray detection assuming they could host a BH (Sect. 4.1, black-filled symbols). We also look for differences with estimates from the \mbh-$\sigma$ relation: the one described by \hyperlink{vdB16}{vdB16} (colour-filled symbols) and the one by \hypertarget{G20}{\citealt{Greene2020}}  (\hyperlink{G20}{G20} hereafter; open symbols). This figure shows that the BH mass estimates using the correlation of \hyperlink{vdB16}{vdB16} are those with the largest differences, followed by the \hyperlink{G20}{G20} ones. This means that the BH mass expected for these galaxies is much smaller than what it is measured. Instead, those estimated from the X-ray luminosities tend to give smaller differences, being in some cases almost the same as the measured value.

There are two possible explanations to account for such differences between the measured and the estimated BH masses. On one hand, large BH masses are expected if the objects are formed via external process. The progenitor, which is a larger and more massive galaxy, will already host a more massive BH. It is expected that as it loses most of its stellar mass it will leave the galaxy's nucleus 'naked' with an untouched BH in the center. This makes the galaxy to move diagonally downwards to the left in the mass-size plane of Figure~\ref{figure:mass_size}, as it also shrinks in size. The only property that remains virtually unchanged during this process is the velocity dispersion (e.g. \citealt{Bekki2003}, \citealt{Pfeffer2013,Pfeffer2016}). 

Depending on the characteristics of the progenitor galaxy (e.g. size, stellar mass, gas content, morphology), the resulting low-mass compact object will be either a UCD or a cE. For the UCDs, it is highly probable that what is left is the NSC that was residing in the center of the progenitor galaxy. For the lowest mass UCDs, it can purely be the NSC, while for more massive UCDs it is possible that it is the naked NSC plus a diffuse region with part of the progenitor galaxy remains (e.g. \citealt{Evstigneeva2007}; \citealt{Pfeffer2013}). Such progenitor are galaxies of \Mstar$\sim10^{8-10}$ \msun, mostly either a low-mass galaxy or a dwarf elliptical (e.g. \citealt{Mieske2013}; \citealt{Pfeffer2016}). However, the recent discovery of the so-called ultradiffuse galaxies, which are nucleated in many cases, also points them as plausible progenitors (\citealt{Janssens2017, Janssens2019}; \citealt{Ferre-Mateu2018b}). In the case of the cEs, they are commonly the result of stripping even more massive galaxies, of \Mstar$\sim10^{10-11}$ \msun. Such progenitors can be either an extended massive galaxy or a massive but compact one \citep{Ferre-Mateu2021}. 

\begin{figure}
\centering
\includegraphics[scale=0.43]{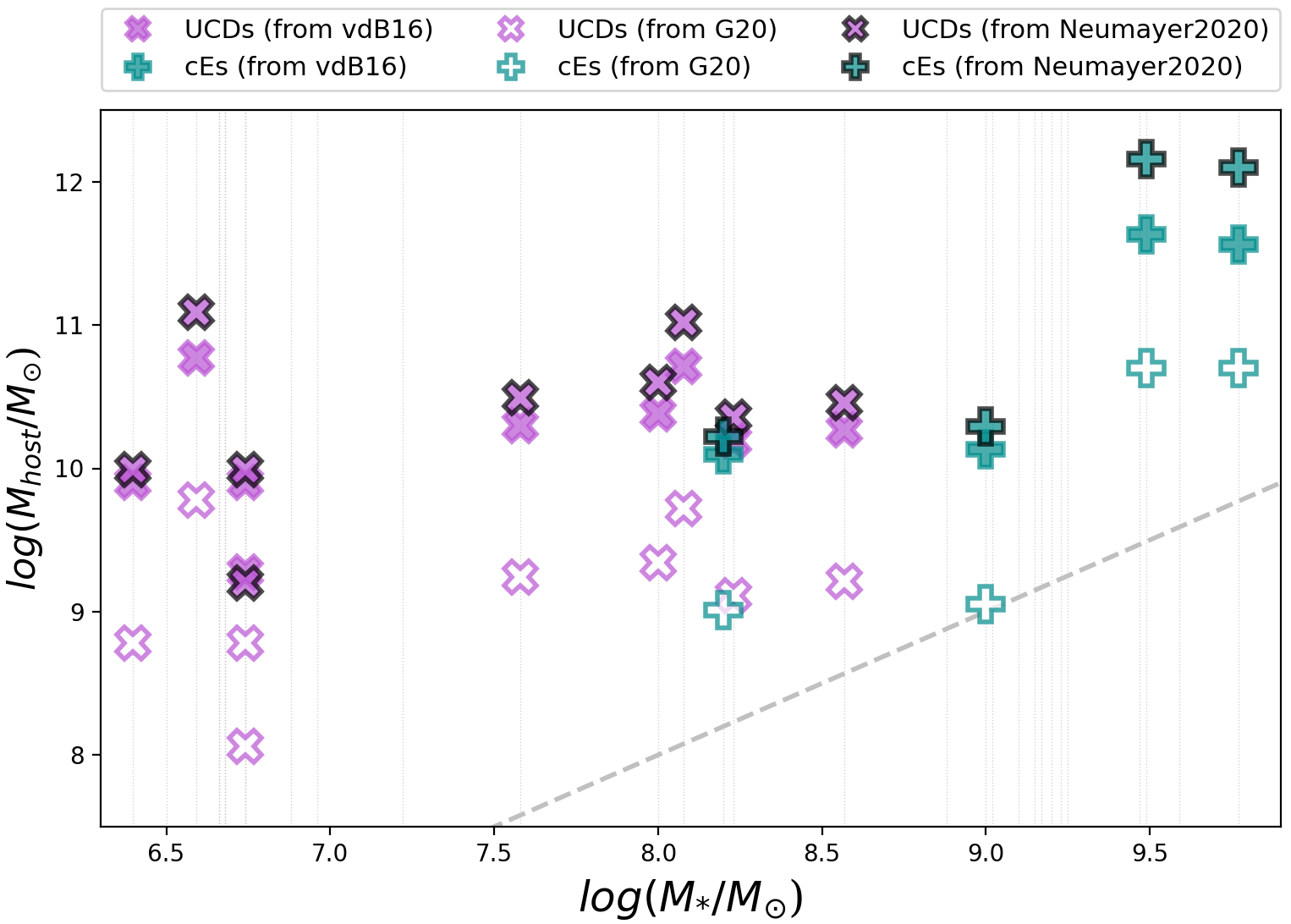}
\caption{Estimated progenitor mass under the assumption that the low-mass compact galaxy is the result of stripping. As in the previous Figure, filled and open colored symbols correspond to using the \mbh-\Mstar relations of vdB16 and G20, respectively. Those colored but outlined in black correspond to using the relation between the \mbh and the NSC as in \citet{Graham2020} and then converting that NSC mass into galaxy mass with the relation from \citet{Neumayer2020}.} 
\label{figure:host_mass}
\end{figure}

One could back-track the BH mass scaling relations to obtain a first approximation of the progenitor stellar mass. If the estimated progenitor stellar mass is similar to the one of the low-mass compact galaxy, then the galaxy could have an intrinsic origin. If the estimated galaxy mass is instead larger than what it is measured, that would be a clear indicator for a stripping origin. This has been previously used to estimate the mass of the possible progenitor for some of the UCDs with direct BH  mass measurements (e.g. \citealt{Mieske2013}; \citealt{Ahn2017}; \citealt{Fahrion2019}; \citealt{Graham2020}). Figure~\ref{figure:host_mass} shows the stellar mass of the low-mass compact galaxies in our sample with direct BH mass measurements compared to the stellar mass estimated for the progenitor following different relations. We use the \mbh-\Mstar relation from \hyperlink{vdB16}{vdB16} (color-filled symbols) and the one from \hyperlink{G20}{G20} (open symbols). Alternatively, we also use the relation between NSC mass and BH mass (e.g. \citealt{Scott2013}; \citealt{Georgiev2016}; \citealt{Graham2020}) and then convert the NSC mass into galaxy stellar mass as in \citet{Neumayer2020}. We use this approach instead of directly assuming that the stellar mass measured from the UCDs is equivalent to the mass of the NSC, because as mentioned before, it could also be the case that it is still surrounded by part of the progenitor galaxy.

We thus use the relation with the BH mass to estimate the NSC masses of the UCDs in our sample, finding NSC masses ranging 10$^{6-7}$\msun\,, which translates into progenitor masses of $\sim10^{10-11}$\msun\,, which is too massive for a UCD progenitor. Similar results are obtained if we use the \hyperlink{vdB16}{vdB16} relation, but those are about one order of magnitude lower if the relation from \hyperlink{G20}{G20} is used. In that case, progenitors of $\sim$10$^{9-10}$\msun\, are found, compatible with the expected type of progenitor for UCDs (\citealt{Bekki2003}, \citealt{Pfeffer2013}). We can now compare our new estimates for the progenitor with those discussed in the literature. For example, for the two UCDs in Virgo from \citet{Ahn2017}, the authors infer progenitor galaxies of $\sim 1.2-1.7 \times 10^{9}$\msun\, following the BH scaling relations of \citet{Saglia2016}. \citet{Seth2014} estimated a bulge mass of $\sim 7\times10^{9}$\msun\, for M60-UCD1 following the scaling relations of \citet{Kormendy2013}, and \citet{Afanasiev2018} infers a progenitor mass of 2$\times10^{9}$\msun\, for Fornax-UCD3. These are all more similar to the BH mass estimates we obtain from the \mbh-$\sigma$ relation of \hyperlink{G20}{G20} than the other estimates.

\begin{figure}
\centering
\includegraphics[scale=0.3]{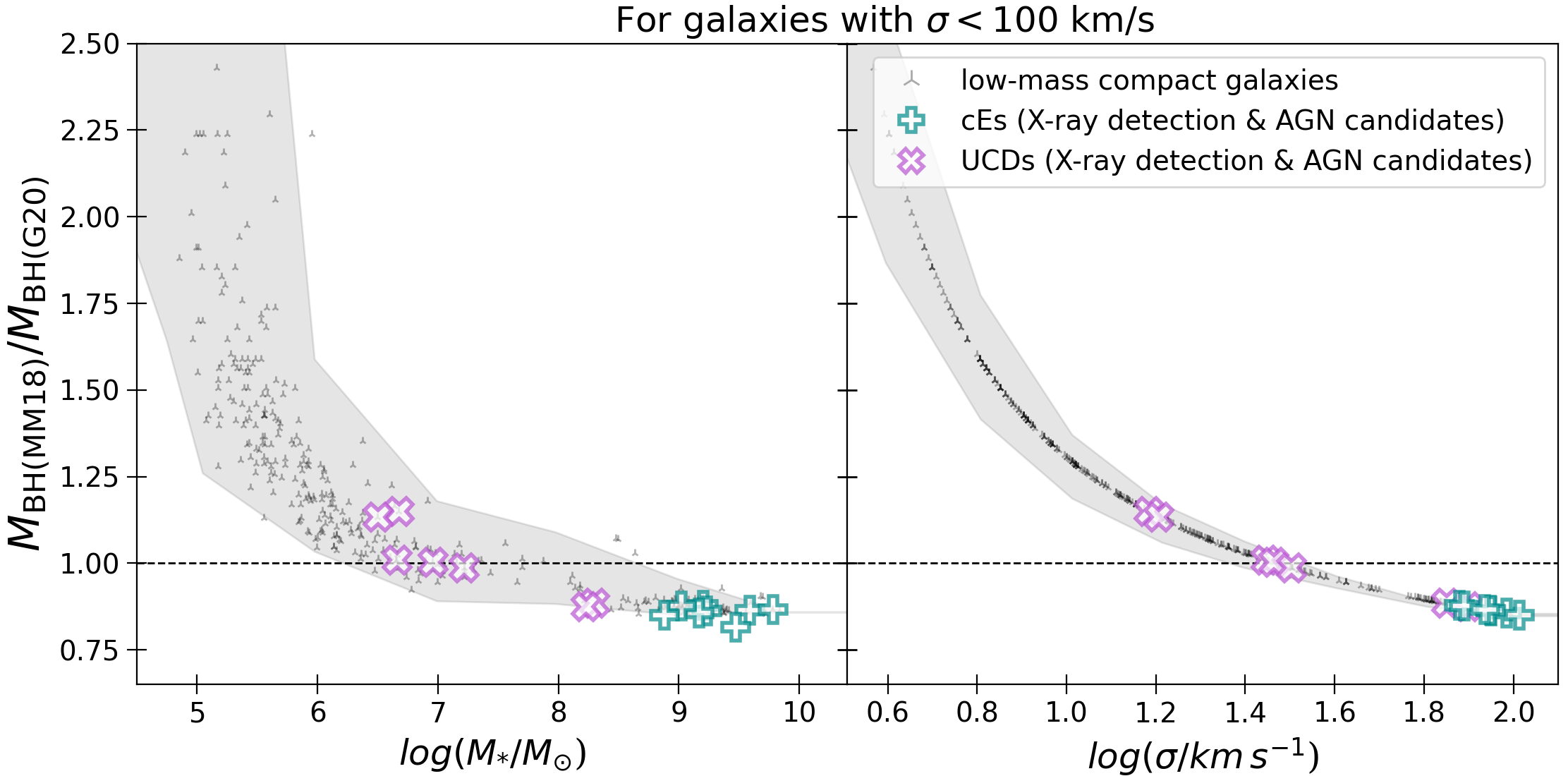}
\caption{Stellar mass (left) and velocity dispersion (right) relations with the ratio between the BH mass from the less steep relation of MM18 and that from the steeper relation of G20 (\mbh$_{\mathrm{(MM18)}}$/\mbh$_{\mathrm{(G20)}}$). Tridents show all the low-mass compact galaxies from this work with a $\sigma <$100\kms and purple and turquoise filled circles correspond to the UCDs and cEs, respectively, with either X-ray detection or that are good AGN candidates. At a fixed $\sigma$, we would expect \mbh$_{\mathrm{(MM18)}}$/\mbh$_{\mathrm{(G20)}} >$1. The 1:1 relation is roughly followed until reaching \Mstar $\lesssim 10^{6}$\msun\, or $\sigma \lesssim$30 \kms.}
\label{figure:smbh_diff}
\end{figure}

\begin{figure*}
\centering
\includegraphics[width=1.00\textwidth]{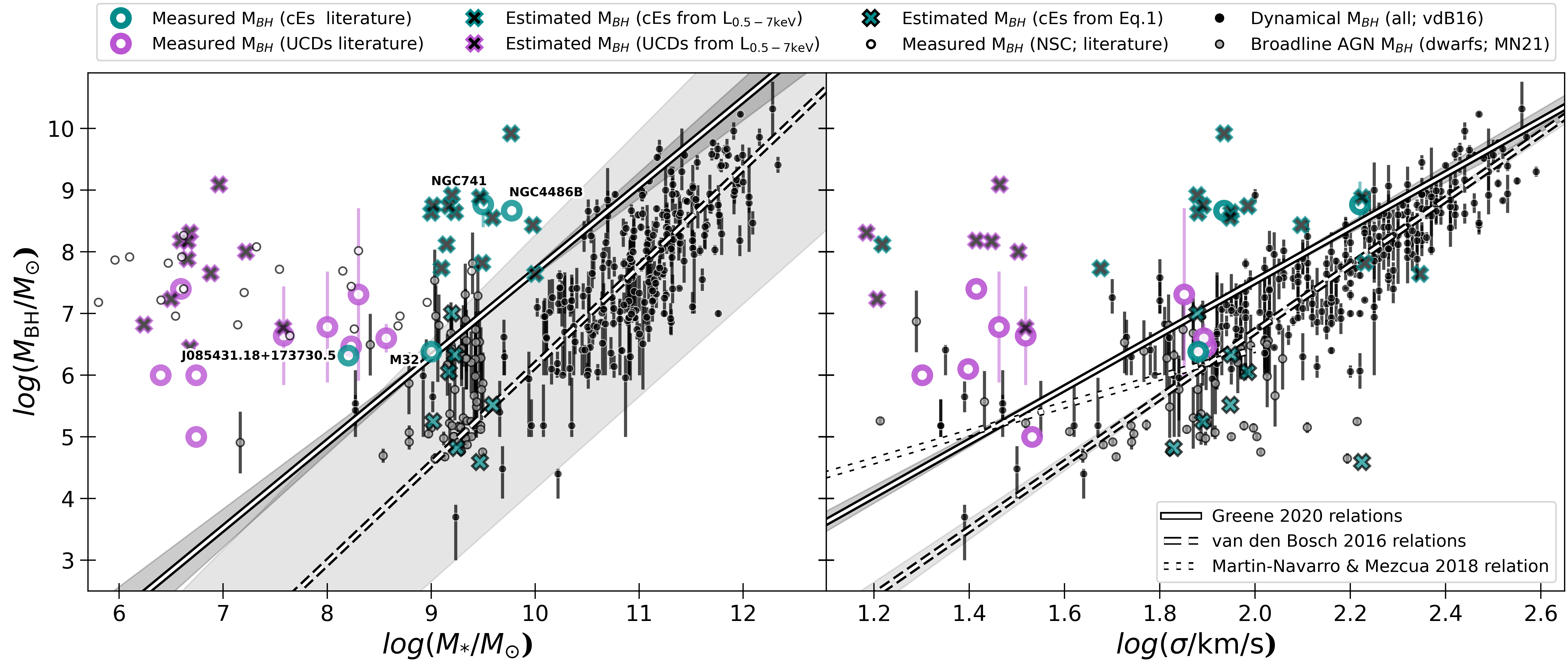}
\caption{Black hole mass relation with galaxy stellar mass (left) and velocity dispersion (right). Measured BH masses from dynamical measurements of vdB16, from AGN broad emission lines from Mart\'in-Navarro et al. 2021 (submitted) and literature NSC (\citealt{Graham2012} and \citealt{Neumayer2020}) are shown in black, grey and white circles, respectively. The published BH masses for UCDs (purple) and cEs (turquoise) are shown in diamonds. Black crosses with colored outline show the BH mass of our sample of low-mass compact galaxies of Table~2 using in section 4.1. turquoise crosses correspond to the estimated values from \citealt{Baron2019} (Eq.~\ref{eq1}). We also show different published BH relations and their intrinsic scatter: G20 (solid white line), vd16 (dashed line) and MM18 (thin dashed line). We remind that the latter should be only considered for galaxies with $\sigma<$100\kms (log$\sigma\sim$2.0\kms), although our results suggest an even lower limit (log$\sigma\sim$1.6\kms).}
\label{figure:smbhs}
\end{figure*}

What is clear from Figure~\ref{figure:host_mass} is that, regardless of the scaling relation employed, low-mass compact galaxies have progenitors larger than expected, suggesting a stripping origin for them. Nonetheless, there are some galaxies around \Mstar$\sim10^{8.5-9}$\msun\ that do not deviate much from the 1:1 relation, which means that these could be good candidates for having an intrinsic origin. In summary, we find that the UCDs in the sample require progenitors that are much more massive than expected by their formation channels. Although their high BH masses can be mostly accounted by the stripping process, something else needs to be considered to match the high values obtained.

The other possibility to explain this effect, although it is still highly debated, is that the scaling relations with BH mass are not universal and therefore can not be extrapolated to the low-mass end. While some works show a particularly tight relation for the \mbh-$\sigma$ that simply becomes more scattered at the low-mass end (e.g. \citealt{Kormendy2013}; \hyperlink{vdB16}{vdB16}; \citealt{Baldassare2020}; \hyperlink{G20}{G20}; \citealt{Shankar2020}), some others suggest that there is a flattening of the \mbh-$\sigma$ relation, which creates a plume at the low-mass end (e.g. \citealt{Greene2006}; \citealt{Mezcua2017}; \hypertarget{MM18}{\citealt{Martin-Navarro2018}}, \hyperlink{MM18}{MM18} hereafter). In the latter, it was suggested that such change in the slope of the \mbh-$\sigma$ relation happens at a $\sigma\sim$100\kms, which roughly corresponds to a stellar mass of 3$\times$10$^{10}$\msun. This is the characteristic mass scale where low-mass compact galaxies separate from the branch of more extended dwarf galaxies (e.g. \citealt{Baldry2012}). This means that, for $\sigma\lesssim$100\kms, the estimated BH mass should be \textit{higher} when using the \hyperlink{MM18}{MM18} relation than a universal one like that of e.g. \hyperlink{G20}{G20} or \hyperlink{vdB16}{vdb16}. 

However, we note that the differences between the measured and expected BH masses happen regardless of the type of \mbh-$\sigma$ relation used to estimate them (e.g. universal \textit{vs} flattened). To further investigate on the possible effect of using one relation or another, Figure \ref{figure:smbh_diff} shows the ratio between the estimated BH mass for the low-mass compact galaxies of our sample (limited to $\sigma <$100\kms) using the BH mass scaling relation of \hyperlink{G20}{G20} and the one from \hyperlink{MM18}{MM18} (the reader can also refer to Figure~\ref{figure:smbhs} for a visualization of such relations). One first thing that should be noted is the difference between the left and right panel uncertainties. While it is largely assumed that the \mbh-$\sigma$ relation is extremely tight, the scatter for the \mbh-\Mstar relation is intrinsically larger, in particular at lower stellar masses. One thing we have not accounted for in our estimates of the previous sections are the proper handling of the uncertainties. This is because, as highlighted in several places, our aim is not to provide exact values but provide averaged estimates to test the hypothesis of a scatter+flattening at the low-mass end. For this reason, we have approximated all our estimates with 0.1\,dex accuracy at most. The left panel, which presents the dependence with the stellar mass, shows that for masses log\Mstar$\gtrsim$6.5 \msun\, it does not make much difference which scaling relation is used, with galaxies being around the 1:1 relation within the intrinsic scatter. This is also seen in the right panel for the velocity dispersion, where it becomes relevant at log$\sigma\lesssim$1.4 \kms (or $\sigma\lesssim$30\kms). Therefore, we find that the impact of using a universal versus a flattened relation is only relevant when going to really low masses. \\

Lastly, we study the scaling relations of BH mass with the velocity dispersion and the stellar mass of galaxies ranging from the high-mass to the low-mass end and including low-mass compact galaxies for the first time, as shown in Figure \ref{figure:smbhs}. Different relations with $M_\mathrm{BH}$ are shown: the one from \hyperlink{G20}{G20} as a solid black and white line, the one from \hyperlink{vdB16}{vdB16} as dashed black and white, and the one from \hyperlink{MM18}{MM18} (up to $\sigma$=100\kms; left panel only). The \hyperlink{vdB16}{vdB16} and \hyperlink{G20}{G20} relations have similar slopes, in particular for the stellar mass panel, but are only compatible at the highest-mass end of the \mbh$-\sigma$ relation. Black dots correspond to the dynamical BH masses from \hyperlink{vdB16}{vdB16}, grey dots are the sample dwarfs with \mbh measured from the broad emission lines of their AGN (MN21), and white dots are a sample of NSCs from \citet{Graham2009} and \citet{Neumayer2020}. Turquoise and purple symbols are cEs and UCDs, respectively, with the direct \mbh measurements from the literature shown in open circles. We also include the \mbh estimated using the $L_\mathrm{0.5-7 keV}$ (black crosses with colored edge-line) and the \mbh from eq.~\ref{eq1} for the cEs that are SDSS AGN candidates (turquoise filled crosses; see Section \ref{section:results}). 

For the UCDs with \Mstar$\lesssim 10^7$\msun\, the values of the BH mass estimates from the X-ray detections are very similar to the values measured for the BHs in NSCs. However, they are both a couple of orders larger than the few BH measurements of low-mass UCDs available. For more massive UCDs (\Mstar$> 10^7$\msun), we find that these objects share the same location as most NSCs. This would further support the idea that more massive UCDs will be, in most of the cases the remnant NSC of a stripped larger galaxy, while lower mass ones can be the result of stripping non-nucleated dwarfs. In the middle of the UCD mass regime there is VUCD3, the only UCD with values for the measured BH mass, an estimated one from the $L_\mathrm{0.5-7 keV}$ and a BH mass estimated from the BH of the NSC. All the three BH masses match remarkably well. However, the BH mass estimates from the X-ray detection seem to provide larger values for the cEs, with only a couple of them having BH masses in the range of some of the measured ones (e.g. similar to NGC~741 and NGC~4486B). We find that for cEs the estimates from Eq~\ref{eq1} seem to provide more similar values to the \mbh measurements of lower-mass cEs (i.e. M32). These are located in the same region occupied by the dwarfs from MN21. Overall, we find that at stellar masses of log\mbh$\sim9$ \msun\, (or log$\sigma\sim$1.8\,\kms) low-mass compact objects tend to be located around the scaling relations, but as we move to lower masses (log\mbh$\sim7$\msun; log$\sigma\sim$1.5\,\kms) they become clear outliers in both panels of Figure \ref{figure:smbhs}. 

Are they outliers simply because of their stripping origin, contributing to the scatter at the low-mass end? Or are they outliers because there is a flattening of the relations? Considering the stripping effect, which is the expected formation channel for the majority of low-mass compact galaxies, and reverse-tracking it, low-mass galaxies would move upwards by $\sim$1-2 orders of magnitude in stellar mass (as we have seen in Figure~\ref{figure:host_mass}), while remaining almost unchanged in the \mbh$-\sigma$ plane (e.g. \citealt{Pfeffer2013, Pfeffer2016}; \citealt{vanSon2019}). Applying these factors to the UCDs in Figure~\ref{figure:smbhs}, low-mass UCDs are expected to have progenitors that are similar to the dwarf galaxies of MN21 (with \Mstar$\sim10^{9}$\msun\,), compatible with the expected progenitors for the known UCDs with BH masses (e.g. \citealt{Mieske2013}). However, for more massive UCDs, their progenitors would be galaxies with \Mstar$\sim10^{10}$\msun\,, supporting the idea that they can also be formed as the result of stripping these more massive galaxies, leaving in many cases a naked NSC as a result. cEs would instead move towards the low-mass end of the realm of ETGs with \Mstar$\sim10^{10-11}$\msun.

In more detail, for example, NGC~4486B and M32 (labelled in the left panel of Fig.~\ref{figure:smbhs}) are both close to the scaling relations in the velocity dispersion panel, but are outliers in the stellar mass one. This suggests that they are threshed galaxies coming from a more massive galaxy. NGC~741 is a clear outlier in both panels but instead J085431.18+173730.5 falls on top of the scaling relation with the stellar mass. This suggests that for this cE an intrinsic nature is plausible, which is in agreement with the original discovery paper \citet{Paudel2016}. UCDs, instead, are all clear outliers in the stellar mass panel, although some of the more massive ones seem to be close to the \mbh-$\sigma$ relations. These are M60-UCD1, M59cO and M59UCD3 and they have all been previously suggested to be the result of stripping a more massive galaxy, most likely a low-mass ETG given their stellar populations with elevated alpha-enhancements (e.g. \citealt{Seth2014}; \citealt{Ahn2017}; \citealt{Afanasiev2018}). Therefore, those galaxies that fall around the scaling relations in both panels will most likely be of intrinsic origin, while those galaxies that are only outliers in the stellar mass panel are most likely of stripped origin. However, the fact that some of them are outliers in both panels might indicate that stripping alone is not enough to account for the scatter in the relations. To date, most of the works that have investigated the low-mass end of the BH scaling relations have focused in the regime of dwarf galaxies (e.g. \hyperlink{MM18}{MM18}; \citealt{Baldassare2020}; \hyperlink{G20}{G20}; \citealt{Shankar2020}). Here we are showing that, if the low-mass objects in the compact branch (cEs and UCDs) are also included, the possibility for the flattening being real becomes more relevant. 

We can account for this flattening in the plane that is less affected by the stripping effect, the \mbh$-\sigma$. We find that if the flattening is real, it seems to happen at lower masses than previously reported by \hyperlink{MM18}{MM18}. The right panel of Figure \ref{figure:smbhs} shows a flattening around $\sigma\sim$20-40~\kms (log$\sigma\sim1.4-1.6$\,\kms), similar to where the change in the 1:1 relation of Figure \ref{figure:smbh_diff} happens. Paired with the previous results of the extremely large BH masses for UCDs, this could indicate that the flattening in the BH scaling relations is real, but happening at lower masses than previously thought. This would thus suggest an even flatter slope for low-mass galaxies than the one presented in \hyperlink{MM18}{MM18}. Therefore, we suggest that the behaviour of the low-mass galaxies is a combination of a flattening and high scatter due to the stripping nature of some of the low-mass galaxies. 

In order to put tighter constraints on the existence of this flattening, it is crucial to populate that low-mass end with a large number of compact galaxies. Unfortunately, due to their extremely small sizes proper studies of their dynamics are both time consuming and limited by the current instrumental resolution. However, with the superb resolution and sensitivity of incoming instruments such as MAVIS and GRAVITY+ at VLT (\citealt{McDermid2020}; \citealt{Eisenhauer2019}) we envision that many more BH measurements in low-mass compact galaxies will be possible. The location of these measurements in the scaling relations will provide additional information regarding the origin of the low-mass compact galaxies (stripped vs. intrinsic) without the need of other discriminants like the stellar populations and kinematics.

\section{Conclusions}\label{section:conclusions}
We have searched for AGN signatures in a sample of low-mass compact objects (580 UCDs and 357 cEs) to probe the origin of these galaxies. If the low-mass compact galaxy is formed intrinsically we expect the BH mass to fall in the IMBH regime (or very low SMBH), whereas if it comes from stripping it could host either an IMBH (if the progenitor galaxy was a dwarf, like some UCDs) or a SMBH (if the progenitor is a more massive galaxy).  

We have searched for signatures in X-rays in the CSC 2.0 catalog and determined the X-ray luminosity. We find that 7 cEs and 23 UCDs have an X-ray detection, corresponding to a detection rate of 7.8\% and 4.8\% respectively. Only one source has an X-ray luminosity indicative of AGN emission, the cE NGC 741, which is a well-studied AGN at several wavelengths. The remaining sources all have X-ray luminosities of $L_\mathrm{0.5-7 keV}\sim10^{37-39}$erg~s$^{-1}$. While this is consistent with X-ray binaries or ultraluminous X-ray sources, a low level of SMBH accretion cannot be fully discarded. 

We also investigate which low-mass compact galaxies have SDSS spectra available in order to identify AGN based on emission-line diagnostic diagrams. We find that only eight cEs can be considered AGN based on their emission line properties. We note the caveat that UCDs have practically no spectra available in SDSS or they have very low S/N. Additionally, we check whether there are any counterparts in the mid-infrared (in WISE) or radio (FIRST). While none of the WISE detections are compatible with an AGN origin, three of the FIRST galaxies are good AGN candidates. Altogether, this translates into an AGN fraction in low-mass compact galaxies of 1.2\% (11/937), which is similar to that found in dwarf galaxies. It increases to a 1.6\% if those objects with weakly accreting SMBHs from the literature are also included. 

Finally, we have studied the BH-galaxy scaling relations for the first time for galaxies of all stellar masses from both the dwarf and the compact branches. We find that compact galaxies tend to be outliers in all local BH-galaxy scaling relations, with larger \mbh than expected. This effect is emphasized in the stellar-mass plane, while it is milder and sometimes even negligible for some galaxies in the \mbh-$\sigma$ plane. Moreover, from existing BH measurements for some of the low-mass compact galaxies, we have inferred how massive their progenitor could have been following different assumptions. We find that in the majority of cases the required progenitor is always more massive than what is expected, regardless of the relation used. Although all these are all strong indicators in favour of a stripping origin for these objects, it also suggests that a flattening in the scaling relations could be real. In fact, it would alleviate the extreme differences found between the measured BH masses and those estimated from the scaling relations. If the flattening is real, we find that it seems to occur at around $\sigma\,\sim$20-40~\kms (log$\sigma\sim$1.3-1.6~\kms), which is a lower value than the one previously reported of $\sim$100~\kms (log$\sigma\sim$2.0~\kms). 

Our results indicate that those UCDs/cEs that are of intrinsic type (i.e. formed naturally as low-mass, compact galaxies rather than being the result of stripping a more massive galaxy) are excellent targets for follow-up studies aimed at populating the low-mass end of the BH scaling relations and posing final constrains on its shape and the existence of IMBHs. 

\section*{Data Availability}
All data is public through the original works in Table~\ref{tab:numbers}. 

\section*{Acknowledgements}
We thank Soo-Chang Rey and Suk Kim for providing their list of cEs from \citet{Kim2020} and Aaron Romanowsky for insightful discussions. AFM has received financial support through the Postdoctoral Junior Leader Fellowship Programme from 'La Caixa' Banking Foundation (LCF/BQ/LI18/11630007). MM acknowledges support from the Beatriu de Pinos fellowship (2017-BP-00114) and the Ramon y Cajal fellowship (RYC2019-027670-I). AFM and MM both acknowledge support from the RAVET project PID2019-107427GB-C32 from the Spanish Ministry of Science and Innovation.

\bibliographystyle{mnras}
\bibliography{smbh_ces}

\bsp	
\label{lastpage}
\end{document}